\documentclass[preprint,floats,tightenlines,11pt,eqsecnum,showpacs,aps]{revtex4}
\usepackage{amsmath}
\usepackage{graphicx}
\begin{document}
\def\appls{\hbox{$<$\kern-.75em\lower 1.00ex\hbox{$\sim$}}}

%\title{STUDY OF $\pi N \to \pi \pi N$ PROCESSES ON POLARIZED TARGETS II:\\
%SPIN MIXING MECHANISM AND THE DARK MATTER}

%\title{STUDY OF $\pi N \to \pi \pi N$ PROCESSES ON POLARIZED TARGETS II:\\
%SPIN MIXING MECHANISM AND THE QUANTUM INFORMATION\\NATURE OF THE DARK MATTER AND DARK ENERGY}

%\title{STUDY OF $\pi N \to \pi \pi N$ PROCESSES ON POLARIZED TARGETS II:\\
%SPIN MIXING MECHANISM, QUANTUM SPACETIME AND \\THE NATURE OF DARK MATTER AND DARK ENERGY}

%\title{STUDY OF $\pi N \to \pi \pi N$ PROCESSES ON POLARIZED TARGETS II:\\
%SPIN MIXING MECHANISM AND QUANTUM SPACETIME}

%\title{STUDY OF $\pi N \to \pi \pi N$ PROCESSES ON POLARIZED TARGETS II:\\
%SPIN MIXING MECHANISM AND QUANTUM SPACETIME ORIGIN\\ OF DARK MATTER AND DARK ENERGY}

%\title{STUDY OF $\pi N \to \pi \pi N$ PROCESSES ON POLARIZED TARGETS II:\\
%SPIN MIXING MECHANISM AND QUANTUM STRUCTURE OF SPACETIME}

%\title{STUDY OF $\pi N \to \pi \pi N$ PROCESSES ON POLARIZED TARGETS II:\\
%SPIN MIXING MECHANISM, DARK MATTER AND DARK ENERGY}

%\title{STUDY OF $\pi N \to \pi \pi N$ PROCESSES ON POLARIZED TARGETS II:\\
%SPIN MIXING MECHANISM AND THE PREDICTION OF $\rho^0(770)-f_0(980)$ MIXING}

\title{STUDY OF $\pi N \to \pi \pi N$ PROCESSES ON POLARIZED TARGETS II:\\
THE PREDICTION OF $\rho^0(770)-f_0(980)$ SPIN MIXING}

%\title{STUDY OF $\pi N \to \pi \pi N$ PROCESSES ON POLARIZED TARGETS II:\\
%FROM $\rho^0(770)-f_0(980)$ SPIN MIXING TO DARK MATTER}

\author{Miloslav Svec\footnote{electronic address: svec@hep.physics.mcgill.ca}}
\affiliation{Physics Department, Dawson College, Montreal, Quebec, Canada H3Z 1A4}
\date{November 28, 2014}

\begin{abstract}

In Part I. of this work we have presented evidence that the measured relative phases of transversity amplitudes in $\pi N \to \pi \pi N$ processes differ from those predicted by the unitary evolution law. We ascribed this difference  to a non-unitary interaction of the produced final state $\rho_f(S)$ with a universal quantum environment in the physical Universe. This new kind of interaction must be a pure dephasing interaction in order to render the $S$-matrix dynamics of particle scattering processes accessible to experimental observation. If the quantum environment is to be an integral part of the Nature then its dephasing interactions must be fully consistent with the Standard Model. 

In this work we impose on the dephasing interaction the requirements of the conservation of the identities of the final state particles including their four-momenta, Lorentz symmetry, $P$-parity and the conservation of total angular momentum and isospin. From this consistency alone we find that the dephasing interaction must be a dipion spin mixing interaction. The observed amplitudes are a unitary transform of the corresponding $S$-matrix amplitudes. The elements of the spin mixing matrix are forward scattering amplitudes of dipion spin states with recoil nucleon spin states with initial and final dipion spins $J$ and $K=J-1,J,J+1$, respectively. Dipion helicities are conserved in this scattering. These amplitudes are matrix elements of Kraus operators describing the non-unitary interaction with the environment. The theory predicts $\rho^0(770)-f_0(980)$ mixing in the $S$-and $P$-wave amplitudes in $\pi^- p \to \pi^- \pi^+ n$. The predicted moduli and relative phases of the mixed amplitudes are in an excellent qualitative agreement with the experimental results. The mixing of $S$-matrix partial wave amplitudes with different dipion spins to form the observable partial wave amplitudes is a new phenomenon beyond the Standard Model. The spontaneous violation of rotational/Lorentz symmetry observed in the measured $S$- and $P$-wave amplitudes is consistent with the conservation of that symmetry by the $S$-matrix and by the Standard Model lagrangian.
\end{abstract}

\maketitle

\tableofcontents

\newpage
\section{Introduction.}

Following the discovery in 1961 of $\rho$ meson in $\pi N \to \pi \pi N$ reactions, the measurements of forward-backward asymmetry in $\pi^- p \to \pi^- \pi^+n$ suggested the existence of a rho-like resonance in the $S$-wave amplitudes, later referred to as $\sigma(750)$ scalar meson~\cite{hagopian63,islam64,patil64,durand65,baton65}. The CERN measurements of $\pi^- p \to \pi^-\pi^+n$ and $\pi^+ n \to \pi^+ \pi^- p$ on polarized targets in 1970's cofirmed the existence of $\sigma(750)$ scalar meson. Evidence for a narrow $\sigma(750)$ was found in amplitude analyses of $\pi^- p \to \pi^-\pi^+n$ at 17.2 GeV/c~\cite{donohue79,becker79a,becker79b,chabaud83,rybicki85,svec92c,svec96,svec97a} and in $\pi^+ n \to \pi^+ \pi^- p$ at 5.98 and 11.85~\cite{svec92c,svec96,svec97a}. Additional evidence for $\sigma(750)$ came from the amplitude analysis of the ITEP data on $\pi^- p \to \pi^- \pi^+ n$ on polarized target at 1.78 GeV/c~\cite{alekseev99}. The $S$-wave amplitudes and intensities from all these analyses are surveyed in Ref.~\cite{svec12d}. These findings were controversial because the measurements of $\pi^- p \to \pi^0 \pi^0 n$ at CERN in 1972 found no evidence for a rho-like meson in the $S$-wave amplitudes~\cite{apel72}. In 2001, E852 Collaboration at BNL reported high statistics measurements $\pi^- p \to \pi^0 \pi^0 n$ at 18.3 GeV/c~\cite{gunter01}. Once again, there was no evidence for a rho-like resonance in the $\pi^0 \pi^0$ $S$-wave amplitudes. 

The resolution of the puzzle of $\sigma(750)$ resonance came from our recent high resolution amplitude analysis of the CERN data on $\pi^- p \to \pi^- \pi^+ n$ for dipion masses 580-1080 MeV at low momentum transfers~\cite{svec12a} where $S$- and $P$-wave amplitudes dominate. The analysis shows that the rho-like resonance in the $S$-wave transversity amplitudes arises entirely from the contribution of the $\rho^0(770)$ resonance to the data component in the expression for the moduli of the $S$-wave amplitudes. The $S$-wave transversity amplitudes are nearly either in phase or 180$^\circ$ out of phase with the resonating $P$-wave transversity amplitudes. These facts allow us to identify $\sigma(750)$ with $\rho^0(770)$. There is a pronounced dip at 980 MeV in the moduli of $P$-wave amplitudes 
$|L_d|^2$ that is identified with $f_0(980)$ resonance leading to $\rho^0(770)-f_0(980)$ mixing in both $S$- and $P$-wave amplitudes. A model independent determination of helicity amplitudes from the transversity amplitudes shows a $\rho^0(770)$ peak in the $S$-wave single flip helicity amplitude $|S_1|^2$~\cite{svec12a}. The phase of the amplitude $S_1$ is close to the phase of $\rho^0(770)$ Breit-Wigner amplitude~\cite{svec12a}.

Identifying $\sigma(750)$ with $\rho^0(770)$ explains why there is no rho-like resonance observed in $\pi^- p \to \pi^0 \pi^0 n$ since there is no $P$-wave amplitude in this process. However, the $\rho^0(770)-f_0(980)$ mixing appears to violate rotational and Lorentz symmetry. If this symmetry were violated in strong interactions we should observe a dependence of $\rho^0(770)$ width $\Gamma_\rho$ on its helicity $\lambda$. In our analysis~\cite{svec12a} we demonstrate the independence of $\Gamma_\rho$ on the helicity confirming the rotational and Lorentz symmetry of strong interactions. This suggests that the observed $\rho^0(770)-f_0(980)$ mixing arises from a new kind of interaction independently involved in the pion creation process.

The observed $\rho^0(770)-f_0(980)$ mixing can be naturally explained when we assume that the produced partial waves are not isolated spin states but interact with a common environment. We can see how this might work using the following mechanical analogy. Consider a pendulum of mass $m$ and length $L$ oscillating with a natural frequency $\omega$, and an ensemble of several other penduli with various masses $m_i$ and lengths $L_i$. When the other penduli are isolated from the oscillating pendulum they do not oscillate. However, when all the penduli are attached to a common rod all penduli begin to oscillate with the same resonant frequency although with different amplitudes. The interaction of the penduli with a common environment - the rod - allows a resonance from one pendulum to "leak" into the other penduli which have different natural frequencies. Similarly we can imagine that the partial waves produced in $\pi^- p \to \pi^- \pi^+ n$ process interact with a quantum environment which allows resonances from one partial wave amplitude to "leak" into other partial wave amplitudes. Then the pion creation process is no longer an isolated system but behaves as an open quantum system described by a non-unitary dynamics. A non-unitary evolution law must evolve the produced final state $\rho_f(S)$ into an observed final state $\rho_f(O)$ with new spin mixing partial wave amplitudes. 
 
But how do we know that such quantum environment actually exists? And why its interaction with the $\pi^- p \to \pi^- \pi^+ n $ process should lead to the $\rho^0(770)-f_0(980)$ mixing?

The basic assumption of the $S$-matrix theory is that the Minkowski spacetime is empty. It contains no environment with which particle scattering processes could interact. In Part I. of this work~\cite{svec13a} we put forward a hypothesis that the physical space is not empty but contains a universal quantum environment that can interact with some particle scattering processes. The new interaction cannot change the identity and the four-momenta of the final state particles produced by the $S$-matrix dynamics in order to leave the particle scattering dynamics open to the experimental observation. This means the new interaction must be a pure dephasing interaction that modifies phases of the amplitudes and nothing else. This is a new kind of interaction outside of the Standard Model. 

To test our hypothesis we used unitarity of the $S$-matrix to obtain information on the relative phases of the measured partial wave transversity amplitudes $U^J_{\lambda \tau}$ and $N^J_{\lambda \tau}$. Unitary evolution law evolves pure initial states into pure final states. This fact implies that the relative phases are constrained to $0, \pm\pi, \pm 2\pi, ...$ in a disagreement with the measured phases in all analyses of $\pi N \to \pi \pi N$ processes. An amplitude analysis of the $S$- and $P$-wave subsystem in $\pi^- p \to \pi^- \pi^+ n$ assuming these unitary phases is in disagreement with the CERN data on polarized target at least at $5\sigma$ level. We conclude that the CERN measurements of $\pi^- p \to \pi^- \pi^+ n$ and $\pi^+ n \to \pi^+ \pi^- p$ on polarized targets support the hypothesis of the existence of the quantum environment and its dephasing interaction with some particle scattering processes.

If the quantum environment is to be an integral part of the Nature then its dephasing interactions with particle scattering processes cannot contradict Standard Model. The two aspects of the Nature must be mutually consistent. What are the cosequences of this consistency? 

The dephasing interaction evolves the produced $S$-matrix final state $\rho_f(S)$ into an observed final state $\rho_f(O)$. The non-unitary evolution law is described by Kraus representation. In this paper we show that the consistency of the pure dephasing interaction with the Standard Model in $\pi N \to \pi \pi N$ processes requires that it be a dipion spin mixing interaction. In what we call spin mixing mechanism the observed partial wave amplitudes are a unitary transform of the corresponding $S$-matrix partial wave amplitudes. The elements of the spin mixing matrix are forward scattering amplitudes of the partial waves with initial and final dipion spins $J$ and $K=J-1,J,J+1$, respectively. Helicities are conserved in this scattering. These amplitudes are matrix elements of Kraus operators describing the non-unitary interaction with the environment. The theory predicts $\rho^0(770)-f_0(980)$ mixing in the $S$-and $P$-wave amplitudes in $\pi^- p \to \pi^- \pi^+ n$. The predicted moduli and relative phases of the mixed amplitudes are in an excellent qualitative agreement with the experimental results~\cite{svec12a,svec12d}. The theory correctly predicts the absence of $f_2(1270)-f_0(1370)$ mixing. The spin mixing of $S$-matrix partial wave amplitudes to form new observable partial wave amplitudes is a new phenomenon outside of Standard Model and represents a genuine new physics. 

We discuss Kraus representation and Kraus operators in Section II. The non-unitary evolution of the produced final state $\rho_f(S)$ into an observed final state $\rho_f(O)$ is introduced in Section III. In Section IV. we use the purity of the dephasing interaction to derive the form of spin mixing mechanism and define Kraus partial wave helicity amplitudes which are mixtures of the $S$-matrix partial wave amplitudes. In Section V. the requirements of Lorentz symmetry, conservation of total angular momentum, $P$-parity and total isospin impose consistency constraints on the spin mixing mechanism and Kraus helicity amplitudes. In Section VI. self-consistency requirements constrain the spin mixing mechanism to a simple unitary transform of the $S$-matrix amplitudes. In Section VII. we relate the observed effective amplitudes to the Kraus amplitudes and demonstrate the conservation of probability by the dephasing intaraction. 

In Section VIII. we discuss resonance mixing in $\pi^- p \to \pi^- \pi^+ n$ below 1400 MeV. We use spin mixing mechanism for the $S$- and $P$-waves to present qualitative predictions for the moduli and relative phases of the spin mixing $S$- and $P$-wave amplitudes below 1080 MeV which we compare with two analyses of experimental data. Section IX. reviews the determination of the dimension $M$ of the Hilbert space of the environment. In Section X. we identify the decoherence free and decohering amplitudes below 1400 MeV in $\pi^- p \to \pi^- \pi^+ n$ and briefly discuss our new analysis using spin mixing mechanism~\cite{svec14a}. In Section XI. we use Coleman-Mandula No-Go Theorem to show that the dimension $M>1$ and that the amplitudes in the decoherence free subspace must mix spins while the decohering amplitudes cannot mix spins to protect the Lorentz symmetry of the $S$-matrix. The spontaneous violation of the Lorentz symmetry observed in the measured amplitudes is thus consistent with Lorentz symmetry of the Standard Model. We conclude in Section XII. with a summary and a brief discussion of the physical nature of the quantum environment.

%\newpage
\section{Non-unitary evolution of open quantum systems.}

The interaction of an open quantum system $S_i$ with a quantum environment $E$ is assumed to be described by a unitary operator $U$. The initial state $\rho_i(S_i,E)$ of the combined system is prepared in a separable state $\rho_i(S_i,E)=\rho_i(S_i) \otimes \rho_i(E)$. The interaction evolves $\rho_i(S_i,E)$ into a final state~\cite{kraus71,kraus83,nielsen00,bengtsson06}
\begin{equation}
\rho_f(S_f,E)=U \rho_i(S_i,E) U^+ 
\end{equation}
Unitary evolutions are entangling quantum operations and the joint state $\rho_f(S_f,E)$ is not a separable but an entangled state of the final system $S_f$ and the environment $E$. The observer can perform measurements on the systems $S_i$ and $S_f$ but cannot perform direct measurements on the environment $E$. After the transformation $U$, the system $S_f$ no longer interacts with the environment $E$. The quantum state of system $S_f$ is then fully described by reduced density matrix
\begin{equation}
\rho_f(S_f)=Tr_E (\rho_f(S_f,E)) 
\end{equation}
in a sense that we can calculate average values $<\hat{O}>=Tr(\hat{O} \rho_f(S_f))$ of any observable $\hat{O}$. The trace in (2.2) is over the interacting degrees of freedom of the environment $|e_\ell>, \ell=1,M$. In general, the reduced state $\rho_f(S_f)$ is no longer related to the initial state $\rho_i(S_i)$ by a unitary transformation $\rho_f=S\rho_i S^+$. Instead it is given by Kraus representation~\cite{kraus71,kraus83,nielsen00,bengtsson06} 
\begin{equation}
\rho_f(S_f)=\sum \limits_{\ell=1}^M A_\ell \rho_i(S_i) A_\ell^+
\end{equation}
where we assumed that $\rho_i(E)=|e_0><e_0|$ is a pure state. The evolution operators  $A_\ell=<e_\ell|U|e_0>$ are called Kraus operators. They are acting on the Hilbert spaces $H(S_i)$ and $H(S_f)$ of the systems $S_i$ and $S_f$, respectively, and can be unitary or non-unitary. For trace preserving maps the Kraus operators must satisfy completness relation
\begin{equation}
\sum \limits_{\ell=1}^M A_\ell^+ A_\ell = I
\end{equation}
To be physically meaningful, a linear map $\rho(S_i) \to  \rho(S_f)=\text{\it{\$}} \rho(S_i)$ must be completely positive in order to preserve the positivity of all probabilities. A linear map $\text{\it{\$}}$ is completely positive if and only if it has the form of the Kraus 
representation~\cite{kraus71,kraus83,nielsen00,bengtsson06}. The operator $\text{\it{\$}}$ is also called superscattering operator. 

The quantum states $|e_\ell>$ of the environment form an orthonormal basis in Hilbert space $H(E)$ with a finite dimension~\cite{nielsen00}
\begin{equation}
M=\dim H(E) \leq \dim H(S_i) \dim H(S_f)
\end{equation}
When the initial state of the environment is a mixed state 
\begin{equation}
\rho_i(E)=\sum \limits_{m,n=1}^M p_{mn}|e_m><e_n|
\end{equation}
the diagonal elements $p_{mm} \geq 0$ and
\begin{equation}
Tr\rho_i(E)=\sum \limits_{m=1}^M p_{mm}=1
\end{equation}
The reduced density matrix (2.2) then takes the form
\begin{equation}
\rho_f(S_f)=\sum \limits_{\ell=1}^M \sum \limits_{m,n=1}^M p_{mn} A_{\ell m} \rho_i(S_i) A^+_{\ell n}
\end{equation}
where $A_{\ell m}=<e_\ell |U|e_m>$. In order for the reduced density matrix (2.8) to be completely positive it must acquire the form of the Kraus representation (2.3). For arbitrary states $\rho_i(E)$ this will happen if and only if the unitary evolution operator $U$ conserves the quantum numbers of the quantum states $|e_\ell>$ 
\begin{equation}
A_{\ell m}=A_\ell \delta_{\ell m}
\end{equation}
where $A_\ell = <e_\ell |U|e_\ell>$. Then (2.8) reads
\begin{equation}
\rho_f(S_f)=\sum \limits_{\ell=1}^M p_{\ell \ell} A_\ell \rho_i(S_i) A_\ell^+= 
\sum \limits_{\ell=1}^M p_{\ell \ell}\rho_f(S_f;\ell)
\end{equation}
With the replacement $\sqrt{p_{\ell \ell}} A_\ell \to A_\ell$ we recover the form (2.3).  We shall refer to $\rho_f(S_f;\ell)$ as Kraus density matrices. The final state $\rho_f(S_f)$ is a mixed state even when the initial state $\rho_i(S_i)$ is a pure state. For trace preserving maps the Kraus operators $A_\ell$ satisfy a completness relation similar to (2.4)
\begin{equation}
\sum \limits_{\ell=1}^M  p_{\ell\ell}A_\ell^+ A_\ell=I
\end{equation}
A trace preserving Kraus representation is called bistochastic if it leaves invariant the maximally mixed state of the system 
\begin{equation}
\rho_i(S)={1\over{N}} \sum \limits_{n=1}^N |n><n|.
\end{equation}
The Kraus representation (2.10) is a bistochastic map when the Kraus operators $A_\ell$ are unitary. In that case the map (2.10) is referred to as a random external field~\cite{bengtsson06}.

When $S_f=S_i=S$ the effect of the non-unitary interaction with the environment described by Kraus representation is a modification (dephasing) of the phases of the quantum state $\rho_i(S)$. In dissipative dephasing interactions the system $S$ and the environment $E$ exchange energy-momentum. There is no exchange of energy-momentum or angular momentum in pure dephasing interactions which effect only a change of phases. 

A very important concept of decoherence free subspace was introduced in Ref.~\cite{lidar99}. It describes a decoupling of a subsystem $S'$ of the system $S$ from the environment. In this case all Kraus operators are equal to a unitary operator $A_\ell=A$ and the Kraus representation is reduced to a unitary evolution law for that subsystem $\rho_f(S')=A\rho_i(S')A^+$.

\newpage
\section{Non-unitary evolution in particle scattering.}

\subsection{Evolution of produced states $\rho_f(S)$ into the observed states $\rho_f(O)$}

Unitary $S$-matrix gives rise to a unitary evolution law
\begin{equation}
\rho(S)=S\rho_iS^+
\end{equation}
The initial state has a general form 
\begin{equation}
\rho_i=\sum \limits_{\nu,\nu'} (\rho_i)_{\nu,\nu'} 
|p_i,\nu;\gamma_i><p_i,\nu';\gamma_i|
\end{equation}
where $p_i,\nu,\gamma_i$ are the initial four-momenta, helicities and quantum numbers of the state. In the following we suppress the labels $p_i$ and $\gamma_i$. The completness relation reads
\begin{equation}
\sum \limits_f \sum \limits_{\xi_f} \int d\Phi_f |p_f,\xi_f,\gamma_f><p_f,\xi_f,\gamma_f|=I
\end{equation}
The first sum in (3.3) is over all allowed final states $f$, the second sum is over final state helicities $\xi_f$ and the integration is over the entire phase space of final state momenta $p_f$. The symbol $\gamma_f$ labels the quantum numbers of the state $f$. Using the completness relation the density matrix $\rho(S)$ has an explicit form 
\begin{equation}
\rho(S)=\sum \limits_f \rho_f(S) +\sum \limits_{f'} \sum \limits_{f''\neq f'} \rho_{f'f''}(S)
\end{equation}
where the diagonal terms
\begin{eqnarray}
\rho_f(S) & = & \sum \limits_{\nu,\nu'} \sum \limits_{\xi_f',\xi_f''} \int d\Phi_f' d\Phi_f'' |p_f',\xi_f',\gamma_f><p_f',\xi_f',\gamma_f|S|\nu>\\
\nonumber
  &  &(\rho_i)_{\nu \nu'}<\nu'|S^+|p_f'',\xi_f'',\gamma_f><p_f'',\xi_f'',\gamma_f|
\end{eqnarray}
are the produced $S$-matrix states with final particle states $|f>$ and
\begin{eqnarray}
\rho_{f'f''}(S) & = & \sum \limits_{\nu,\nu'} \sum \limits_{\xi_{f'},\xi_{f''}} \int d\Phi_{f'} d\Phi_{f''}
|p_{f'},\xi_{f'},\gamma_{f'}><p_{f'},\xi_{f'},\gamma_{f'}|S|\nu>\\
\nonumber
 &  &(\rho_i)_{\nu \nu'}<\nu'|S^+|p_{f''},\xi_{f''},\gamma_{f''}><p_{f''},\xi_{f''},\gamma_{f''}|
\end{eqnarray}
are the off-diagonal terms of $\rho(S)$.

We assume that particle scattering processes are not isolated events in the Universe as implied by the unitary evolution law but behave as open quantum systems interacting with a universal quantum environment of quantum states $\rho(E)$. If the quantum environment and its interaction with particle scattering processes are to be an integral part of the Nature then they must be fully consistent with the Standard Model. This is possible if this new kind of particle interaction does not mix with the fundamental interactions of the Standard Model and thus does not participate in the scattering or decay process itself. Then the interaction with the quantum states $\rho(E)$ can only evolve the state $\rho(S)$ produced by the $S$-matrix dynamics into a new observed state $\rho(O)$. The observed state $\rho(O)$ is described by the non-unitary Kraus representation
\begin{equation}
\rho(O)=\sum \limits_{\ell=1}^M p_{\ell\ell}V_\ell\rho(S)V_\ell^+=\sum \limits_f \rho_f(O) +\sum \limits_{f'} \sum \limits_{f''\neq f'} \rho_{f'f''}(O)
\end{equation}
where
\begin{eqnarray}
\rho_f(O) & = & \sum \limits_{\ell=1}^M p_{\ell\ell}V_\ell\rho_f(S)V_\ell^+\\
\rho_{f'f''}(O) & = & \sum \limits_{\ell=1}^M p_{\ell\ell}V_\ell
\rho_{f'f''}(S)V_\ell^+
\end{eqnarray}
It is the observed states $\rho_f(O)$ with final particle states $|f>$ which are the object of experimental measurements.

The states $\rho(E)$ and thus the probabilities $p_{\ell\ell}$ are Poincare invariant. The Kraus operators $V_\ell$ are unitary to preserve in (3.8) the completness relations (3.3). The observed states $\rho_f(O)$ must communicate the information about the scattering dynamics encoded in the $S$-matrix state $\rho_f(S)$. The Kraus operators therefore must preserve the identities of all final state particles of the produced state $\rho_f(S)$ including the four-momenta of individual particles. This means that we need to concern ourselves only with the projective measurements of the states $\rho_f(O)$ as the projections of $\rho_{f'f''}(O)$ vanish.   

There is no exchange of four-momentum and angular momentum between the environment and the produced state $\rho_f(S)$. The total four-momentum and total angular momentum are conserved by the Kraus operators. The interaction of  the produced state $\rho_f(S)$ with the quantum environment is a non-dissipative pure dephasing interaction. The consistency of this interaction with the symmetries and conservation laws of the Standard Model imposes constraints on the Kraus operators and their matrix elements.

\subsection{Experimental signature of the dephasing interaction}

How do we observe the changes of the $S$-matrix amplitudes in the observed amplitudes from the measurements of the density matrix $\rho_f(O)$? To answer this question consider three complex functions $A,B,C$. Their relative phases $\Phi(AB^*)=\Phi(A)-\Phi(B),\Phi(AC^*)=\Phi(A)-\Phi(C)$ and $\Phi(CB^*)=\Phi(C)-\Phi(B)$ satisfy the phase condition
\begin{equation}
\Phi(AB^*)=\Phi(AC^*) + \Phi(CB^*)
\end{equation}
and the equivalent cosine condition
\begin{equation}
\cos^2\Phi(AB^*)+\cos^2\Phi(AC^*)+\cos^2\Phi(CB^*)-
2\cos\Phi(AB^*)\cos\Phi(AC^*)\cos\Phi(CB^*)=1
\end{equation}
$S$-matrix amplitudes $A(S),B(S),C(S)$ are complex valued functions that satisfy the cosine condition (3.11). Another fundamental property of the $S$-matrix amplitudes is that they do not mix resonances of different spins $J_R$ and $J_R'$ in the bilinear terms $|A(S)|^2,|B(S)|^2$ and $Re(A(S)B(S)^*)$. 

To be physically meaningfull the observed density matrix $\rho_f(O)$ and the Kraus densities $\rho_f(O,\ell)=V_\ell \rho_f(S) V^+_\ell$ must have the same bilinear structure as the $S$-matrix density $\rho_f(S)$. Then the measured bilinears $|A(O)|^2,|B(O)|^2$ and $Re(A(O)B(O)^*)$ carry the same quantum numbers as the corresponding $S$-matrix bilinear terms. From the measured bilinear terms we can determine three cosines $\cos \Phi(A(O)B(O)), \cos \Phi(A(O)C(O)), \cos \Phi(C(O)B(O))$. Since the observed amplitudes are not $S$-matrix amplitudes, these cosines will, in general, violate the cosine condition. For the same reason in some processes certain bilinear terms will exhibit spin mixing of resonances. 

Only in the case of a decoherence free subspace involving at least three amplitudes will the observed cosines satisfy the cosine condition (3.11). If there is no spin mixing observed in these amplitudes then they can be identified with $S$-matrix amplitudes. This is the case of all two-body scattering or decay processes. These processes as well as free single particles thus do not interact with the quantum environment.

\section{Dephasing interaction and the spin mixing mechanism\\ in $\pi N \to \pi \pi N$.}

\subsection{$S$-matrix final state $\rho_f(S)$ in $\pi N \to \pi \pi N$}

The pion creation process $\pi_a N_b \to \pi_1 \pi_2 N_d$ is the very simplest particle scattering process where we can observe non-unitary evolution and the spin mixing interaction. With four-momenta $p_a+p_b \to p_1+p_2+p_d$ the initial and final particle states of the process are described by state vectors $|p_a p_b;0\nu>$ and $|p_c p_d;\theta \phi; \chi>$ where $p_c=p_1+p_2$ and $\nu$ and $\chi$ are the helicities of the target and recoil nucleon, respectively. The direction of $\pi_1$ in dipion center-of-mass system is descibed by $\theta, \phi$. Target nucleon spin density matrix is $\rho_b(\vec{P})$ where $\vec{P}$ is target polarization. The produced final state is given by the $S$-matrix unitary evolution law
\begin{equation} 
\rho_f(S)  = 
\int d\Phi_3 d\Phi_3' \sum \limits_{\xi \xi'} 
R(S,\vec{P})_{\xi \xi'}|p_cp_d;\theta \phi,\xi><p_c'p_d';\theta' \phi',\xi'|
\end{equation}
where
\begin{equation}
R(S,\vec{P})_{\xi \xi'} = \sum \limits_{\nu \nu'}
S_{\xi,0\nu}(p_cp_d,\theta\phi)\rho_b(\vec{P})_{\nu\nu'}
S_{\xi',0\nu'}^*(p_c'p_d',\theta'\phi')
\end{equation}
and where we have used the phase space relation~\cite{martin70} 
\begin{equation}
d\Phi_3=
{d^3 \vec{p}_1 \over{2E_1}}{d^3 \vec{p}_2 \over{2E_2}}{d^3 \vec{p}_d \over{2E_d}}
={q\over{4m}}d^4p_c d\Omega{d^3 \vec{p}_d \over{2E_d}}=
d\overline{\Phi}_3 d\Omega
\end{equation}
in the completness relation. The angular $S$-matrix amplitudes are matrix elements
\begin{eqnarray}
S_{\xi,0\nu}(p_cp_d,\theta\phi) & = &
<p_cp_d;\theta \phi,\xi|S|p_ap_b,0\nu>\\
\nonumber
S_{\xi',0\nu'}^*(p_c'p_d',\theta'\phi') & = & 
<p_ap_b,0\nu'|S^+|p_c'p_d';\theta' \phi',\xi'>
\end{eqnarray}
Prior to the interaction of the produced final state with the quantum environment there is no measurement and thus no measurement projection of this state to $\rho_f(p_c p_d,\theta \phi; \vec{P})$). The angular state $|p_c p_d;\theta \phi, \xi>$  can be expanded in terms of spherical harmonics~\cite{svec13a}
\begin{equation}
|p_c p_d;\theta \phi, \xi> = \sum \limits_{K=0}^{\infty} \sum \limits_{\mu=-K}^K Y^{K*}_{\mu}(\theta, \phi) |p_c p_d;K \mu, \xi>
\end{equation}
where $K$ and $\mu$ are the dipion spin and helicity, respectively. Integrating over $d\Omega$ and $d\Omega'$ and using the relation
\begin{equation}
\int d\Omega Y^K_\mu(\Omega)Y^{K'*}_{\mu'}(\Omega)=\delta_{KK'} \delta_{\mu \mu'}
\end{equation}
we find
\begin{equation}
\rho_f(S)=
\int d\overline{\Phi}_3 d\overline{\Phi'}_3
\sum \limits_{K\mu,\xi} \sum \limits_{K'\mu',\xi'}
R(S,\vec{P})^{K{1\over{2}},K'{1\over{2}}}_{\mu \xi,\mu'\xi'}
|p_c p_d;K\mu,\xi><p_c'p_d';;K'\mu',\xi'|
\end{equation}
where the joint spin density matrix elements
\begin{equation}
R(S,\vec{P})^{K{1\over{2}},K'{1\over{2}}}_{\mu \xi,\mu'\xi'}(p_cp_d,p_c'p_d')=
\sum \limits_{\nu,\nu'} S^K_{\mu \xi, 0 \nu}(p_cp_d)
\rho_b(\vec{P})_{\nu\nu'}S^{{K'}*}_{\mu' \xi', 0 \nu'}(p_c'p_d')
\end{equation}
The partial wave $S$-matrix amplitudes are matrix elements
\begin{eqnarray}
S^K_{\mu \xi, 0 \nu}(p_cp_d) & = & <p_cp_d;K \mu,\xi|S|p_ap_b,0\nu>\\
\nonumber
S^{{K'}*}_{\mu' \xi', 0 \nu'}(p_c'p_d') & = & <p_ap_b,0\nu'|S^+|p_c'p_d';
K' \mu',\xi'>
\end{eqnarray}
In general, the final state density matrix (4.7) is a mixed state of entangled  dipion and recoil nucleon helicity helicity states $|K\mu>$ and $|\xi>$ with different dipion spins $K$. Equation (4.7) can be written in a physically important form
\begin{equation}
\rho_f(S)=
\int d\overline{\Phi}_3 d\overline{\Phi'}_3
\sum \limits_{\nu \nu'} 
\rho_i(\vec{P})_{\nu \nu'}|\Psi(p_c p_d,0\nu)><\Psi(p_c' p_d',0\nu')|
\end{equation}
where $|\Psi(p_c p_d,0\nu)>$ and $<\Psi(p_c' p_d',0\nu')|$ are superpositions of partial waves $|p_cp_d;K \mu,\xi>$ given by
\begin{eqnarray}
|\Psi(p_c p_d,0\nu)> & = & \sum \limits_{K\mu,\xi} 
S^K_{\mu \xi, 0 \nu}(p_cp_d)|p_cp_d; K\mu,\xi>\\
\nonumber
<\Psi(p_c' p_d',0\nu')| & = & \sum \limits_{K'\mu',\xi'} 
<p_c'p_d'; K'\mu',\xi'|S^{{K'}*}_{\mu' \xi', 0 \nu'}(p_c'p_d')
\end{eqnarray}

\subsection{Spin mixing mechanism}

We now use the Kraus representation (3.7) to determine the nature and the form of the pure dephasing interaction with the quantum environment in $\pi N \to \pi \pi N$ processes. The unitary Kraus operators $V_\ell$ act on the superposions (4.11) of the partial waves $|p_cp_d;K\mu, \xi>$ by rotating the dipion spin states $|K\mu,\xi>$. Applying the completness relation (3.3) from left to $V_\ell |p_cp_d;K\mu,\xi>$ we find 
\begin{eqnarray} 
V_\ell|p_cp_d;K\mu, \xi> & = & \sum \limits_f \sum \limits_{\xi_f} \int d\Phi_f |p_f,\xi_f,\gamma_f><p_f,\xi_f,\gamma_f|V_\ell|p_c p_d;K\mu, \xi>\\
\nonumber
  & = & \sum \limits_\chi \int d\Omega |p_cp_d;\theta \phi,\chi><p_cp_d;\theta \phi,\chi|
V_\ell|p_c p_d;K\mu, \xi>
\end{eqnarray}
since the Kraus operators preserve the four-momenta and the identity of the final state particles. Using the identity
\begin{equation}
\sum \limits_\chi \int d\Omega |\theta \phi,\chi><\theta \phi,\chi|=
\sum \limits_{J\lambda,\chi}|J\lambda,\chi><J\lambda,\chi|=I
\end{equation}
the equation (4.12) then acquires the form
\begin{equation}
V_\ell|p_cp_d;K\mu, \xi> =\sum \limits_{J\lambda,\chi}
<p_cp_d;J\lambda,\chi|V_\ell|p_cp_d;K\mu,\xi>|p_cp_d;J\lambda,\chi>
\end{equation}
This relation shows the interaction of the produced state $\rho_f(S)$ with the environment is a dipion spin mixing interaction. Its effect is to produce a new superposition of partial waves $|p_cp_d;J\lambda,\chi>$
\begin{equation}
|\Psi_\ell(p_c p_d,0\nu)>=V_\ell|\Psi(p_c p_d,0\nu)>=
\sum \limits_{J\lambda,\chi} A^J_{\lambda \chi,0\nu}(p_cp_d;\ell)
|p_cp_d;J\lambda,\chi>
\end{equation}
Here $A^J_{\lambda \chi,0\nu}(p_cp_d;\ell)$ are Kraus partial wave amplitudes
\begin{equation}
A^J_{\lambda \chi,0\nu}(p_cp_d;\ell)=
\sum \limits_{K,\mu} \sum \limits_\xi 
<p_cp_d;J\lambda,\chi|V_\ell|p_cp_d;K\mu,\xi><p_cp_d;K \mu,\xi|S|p_ap_b,0\nu>
\end{equation}
which explicitely show the mixing of $S$-matrix partial wave amplitudes with different dipion spins $K$ and helicities $\mu$. We refer to the relation (4.16) as spin mixing mechanism. Applying the relation (4.13) to the combination 
\begin{equation}
\sum \limits_{J\lambda,\chi} 
|p_cp_d;J\lambda,\chi><p_cp_d;J\lambda,\chi|V_\ell|p_cp_d;K\mu,\xi>
\end{equation}
in (4.15) we transform $|\Psi_\ell(p_c p_d,0\nu)>$ into an angular form
\begin{equation}
|\Psi_\ell(p_c p_d,0\nu)> =
\int d\Omega |p_cp_d;\theta \phi,\chi>A_{\chi,0\nu}(p_cp_d,\theta \phi;\ell)
\end{equation}
where the Kraus angular amplitudes
\begin{equation}
A_{\chi,0\nu}(p_cp_d,\theta \phi;\ell)=\sum \limits_{K \mu,\xi}
<p_cp_d;\theta \phi, \chi|V_\ell|p_cp_d;K\mu,\xi>
<p_cp_d;K \mu,\xi|S|p_ap_b,0\nu>
\end{equation}
Similarly we find for $<\Psi_\ell(p_c' p_d',0\nu')|=<\Psi(p_c' p_d',0\nu')|V^+_\ell$
\begin{equation}
<\Psi_\ell(p_c' p_d',0\nu')| = 
\sum \limits_{J'\lambda',\chi'}
A^{{J'}*}_{\lambda' \chi',0\nu'}(p_c'p_d';\ell)
<p_c'p_d';J'\lambda',\chi'|
\end{equation}
\[
= \int d\Omega' A_{\chi',0\nu'}^*(p_c'p_d',\theta'\phi';\ell)
<p_c'p_d';\theta'\phi',\chi'|
\]
where the Kraus amplitudes
\begin{equation}
A^{{J'}*}_{\lambda' \chi',0\nu'}(p_c'p_d';\ell) = 
\sum \limits_{K'\mu',\xi'}<p_ap_b,0\nu'|S^+|p_c'p_d',K'\mu',\xi'>
<p_c'p_d';K'\mu',\xi'|V^+_\ell|p_c'p_d';J'\lambda',\chi'>
\end{equation}
\[
A^*_{\chi',0\nu'}(p_c'p_d',\theta'\phi';\ell) =
\sum \limits_{K'\mu',\xi'}<p_ap_b,0\nu'|S^+|p_c'p_d',K'\mu',\xi'>
<p_c'p_d';K'\mu',\xi'|V^+_\ell|p_c'p_d';\theta'\phi',\chi'>
\]
The Kraus representation (3.7) then reads
\begin{equation}
\rho_f(O)=\sum \limits_{\ell=1}^M p_{\ell\ell} \rho_f(\vec{P};\ell\ell)
\end{equation}
where
\begin{eqnarray}
\rho_f(\vec{P};\ell\ell) & = &
\int d\Phi_3 d\Phi_3' \sum \limits_{\chi \chi'} \sum \limits_{\nu \nu'}
|p_cp_d;\theta \phi,\chi><p_c'p_d';\theta' \phi',\chi'|\\
\nonumber
 &  & A_{\chi,0\nu}(p_cp_d,\theta \phi;\ell)\rho_b(\vec{P})_{\nu\nu'}
A_{\chi',0\nu'}^*(p_c'p_d',\theta'\phi';\ell)
\end{eqnarray}

The equations (4.23) have a formal resemblance to the equation (4.1) for the $S$-matrix final state $\rho_f(S)$. It is evident from the definitions of Kraus partial wave and angular amplitudes (4.16), (4.19) and (4.21) that these amplitudes cannot be written as matrix elements of a unitary operator in contrast to the $S$-matrix partial wave amplitudes if not all spin states $|K \mu,\xi>$ are allowed and (4.13) does not apply. In that case the Kraus partial wave amplitudes shall violate the unitarity conditions on relative phases of the $S$-matrix partial wave amplitudes~\cite{svec13a}. As a result they are complex valued functions with non-unitary phases. Spins $K$ are resticted by the conditions (5.6).

\subsection{Projective measurement and Kraus helicity amplitudes}

Following a projective measurement of the final state $\rho_f(O)$ into a state with definite four-momenta $p_c,p_d$ and the direction $\theta\phi$ of the pion $\pi_1$, the Kraus representation (4.22) takes the form
\begin{equation}
\rho_f(p_cp_d,\theta \phi,\vec{P})=\sum \limits_{\ell=1}^M p_{\ell\ell}\rho_f(p_cp_d,\theta \phi, \vec{P};\ell\ell)
\end{equation}
Suppressing the four-momenta in the initial and final state vectors $|0 \nu>$ and $|\theta \phi, \chi>$, the Kraus density matrices read
\begin{equation}
\rho_f(\theta \phi, \vec{P};\ell\ell)_{\chi \chi'}=\sum \limits_{\nu \nu'}
A_{\chi,0\nu}(\theta \phi;\ell)\rho_b(\vec{P})_{\nu \nu'}
A_{\chi',0\nu'}(\theta \phi;\ell)
\end{equation}
With $S=I+i(2\pi)^4\delta(P_f-P_i)T$ we have from (4.13)
\begin{equation}
A_{\chi,0\nu}(\theta \phi;\ell)= i(2\pi)^4 \delta(P_f-P_i)
T_{\chi,0\nu}(\theta \phi;\ell)
\end {equation}
where $T_{\chi,0\nu}(\theta \phi;\ell)$ are the new Kraus angular amplitudes. Following formally the steps employed in the case of the $S$-matrix spin formalism~\cite{svec13a} we can redefine the Kraus density matrices (4.25) in terms of new Kraus helicity amplitudes 
\begin{equation}
H_{\chi, 0 \nu}(\theta \phi;\ell)=K(s,m^2)T_{\chi,0\nu}(\theta \phi;\ell)
\end{equation}
The normalization factor
\begin{equation}
K(s,m^2)= \sqrt{{q(m^2)G(s)\over{Flux(s)}}}
\end{equation}
where $m^2=p_c^2$ is the dipion mass, $s$ is the center-of-mass energy squared,  $q(m^2)$ is pion momentum in dipion center-of-mass, $G(s)$ and $Flux(s)$ are the energy dependent part of the phase space and flux, respectively~\cite{svec97a}. The redefined matrices (4.25) then read
\begin{equation}
\rho_f(\theta \phi, \vec{P};\ell\ell)_{\chi \chi'}=
\sum \limits_{\nu \nu'}H_{\chi,0 \nu}(\theta \phi;\ell)\rho_b(\vec{P})_{\nu \nu'} H^*_{\chi', 0 \nu'}(\theta \phi;\ell)
\end{equation}
Angular expansion of Kraus helicity amplitudes
\begin{equation}
H_{\chi,0\nu}(\theta \phi;\ell)=\sum \limits_{J=0}^\infty 
\sum \limits_{\lambda=-J}^J H^J_{\lambda \chi, 0 \nu}(\ell) Y^J_\lambda(\theta,\phi)
\end{equation}
defines Kraus partial wave helicity amplitudes with a definite dipion spin and helicity
\begin{equation}
H^J_{\lambda \chi, 0 \nu}(\ell)=K(s, m^2)T^J_{\lambda\chi,0\nu}(\ell)
\end{equation}

So far we have suppressed isospin labels of the produced pions. In strong interactions pions are considered as identical particles. The ensuing Bose-Einstein statistics for two-pion states requires that $I+J=even$ where $I$ is the total dipion isospin~\cite{martin70}. Only $J=even$ is allowed for identical pions. For different pions $S$-matrix partial waves with $I+J =odd$ vanish while for identical pions amplitudes with $J=odd$ vanish. Labeling the isospin states $I_J$, the spin mixing mechanism for Kraus partial wave helicity amplitudes reads
\begin{equation}
H^J_{\lambda \chi,0\nu}(\ell)=
\sum \limits_{K,\mu} \sum \limits_\xi 
<p_cp_d;J\lambda I_J,\chi|V_\ell|p_cp_d;K\mu I_K,\xi>H^K_{\mu \xi,0\nu}(S)
\end{equation}
where $H^K_{\mu \xi,0\nu}(S)$ are the redefined $S$-matrix partial wave helicity amplitudes~\cite{svec13a}. The isospin of the Kraus partial wave helicity amplitudes is defined to be $I_J$ although for different pions these amplitudes mix isospin as the result of spin mixing. Kraus partial wave amplitudes with $J=odd$ still vanish for identical pions due to the Bose-Einstein statistics for pion spin.

\section{Consistency of the dephasing interaction with the Standard Model.}

\subsection{Lorentz symmetry of the Kraus operators}

We refer to the matrix elements $<p_cp_d;J\lambda I_J,\chi|V_\ell|p_cp_d;K\mu I_K,\xi>$ as dephasing amplitudes. They describe a forward two-body scattering of dipion states and recoil nucleons
\begin{equation}
|p_c,K\mu I_K> + |p_d,\frac{1}{2} \mu> \to |p_c,J\lambda I_J> + |p_d,\frac{1}{2} \chi>  
\end{equation}
with the Kraus operators $V_\ell$ taking on the role of scattering operators.
As in the case of $S$-matrix scattering, there is a priori no restriction on the final spin $J$ and helicity $\lambda$. 

To be consistent with the Standard Model, the dephasing interaction must be Lorentz invariant. That means that all Kraus operators must be Lorentz invariant. The rescattering process of the dipions with the recoil nucleon must thus conserve the total four-momentum and total angular momentum.

The dephasing amplitudes obviously conserve the total four-momentum. They must depend on the Lorentz invariants of the four-momenta $p_c$ and $p_d$, i.e. the
total energy $s$ and dipion mass $m^2$. To examine their conservation of the total angular momentum we go into the center-of-mass system in the initial and final states of the process (4.33) where the dipion has 3-momentum q and direction $\theta_c, \phi_c$. Then~\cite{martin70}
\begin{eqnarray}
|p_cp_d,K\mu I_K,\xi> & = & \sqrt{\frac{4s}{q}} \sum \limits_{j_i m_i} \sqrt{\frac{2j_i+1}{4 \pi}}\mathcal{D}^{j_i}_{m_i \mu'}(\phi_c,\theta_c,-\phi_c)
|q,j_im_i,K\mu I_K,\xi>\\
\nonumber
<p_cp_d,J\lambda I_J,\chi| & = & \sqrt{\frac{4s}{q}} \sum \limits_{j_fm_f} \sqrt{\frac{2j_f+1}{4 \pi}}\mathcal{D}^{j_f*}_{m_f \lambda'}(\phi_c,\theta_c,-\phi_c)
<q,j_fm_f,J\lambda I_J,\chi|
\end{eqnarray}
where $\mu'=\mu-\xi$ and $\lambda'=\lambda-\chi$. The expansion of the dephasing amplitudes defines partial wave amplitudes 
in $<q,j_f m_f,J\lambda I_J,\chi|V_\ell|q,j_im_i,K\mu I_K,\xi>$  which conserve the total angular momentum $j_i=j_f$, $ m_i=m_f$ and which are independent of $m_i$ due to rotational invariance. Then
\begin{equation}
<p_cp_d;J\lambda I_J,\chi|V_\ell|p_cp_d;K\mu I_K,\xi>= 
\delta_{\lambda' \mu'} <q;J\lambda I_J,\chi|V_\ell|q;K\mu I_K,\xi>
\end{equation}
where
\begin{equation}
<q;J\lambda I_J,\chi|V_\ell|q;K\mu I_K,\xi>=
\frac{4s}{q} \sum \limits_{j_i}
\frac{2j_i+1}{4\pi}<q;j_i,J\lambda I_J,\chi|V_\ell|q;j_i,K\mu I_K,\xi>
\end {equation}
We see that the dephasing amplitudes are non-zero only for
\begin{equation}
\lambda-\chi=\mu-\xi
\end{equation}
We can write the conservation of total angular momentum $j_f=j_i$ in the form $L_i+S_i=L_f+S_f$ where $L_i$ and $L_f$ are the initial and final orbital momenta while $S_i$ and $S_f$ are the initial and final total spins. Since the four-momenta of the particles in the initial and final states are equal we have $L_i=L_f$. For total spins we have $S_i=K\pm\frac{1}{2}$ and $S_f=J \pm\frac{1}{2}$. Then $S_i=S_f$ implies that the only possible values of the spin $K$ are
\begin{equation}
K=J-1,J,J+1
\end{equation}

\subsection{Consistency of Kraus helicity amplitudes}

We require that the Kraus helicity amplitudes (4.32) be consistent with the Standard Model, in particular with the conservation of $P$-parity, total four-momentum and total angular momentum, total isospin, and with the conservation of probability imposed by the $S$-matrix. These requirements will impose constraints on the matrix elements of the Kraus operators that ensure the dipion spin mixing interaction is a pure dephasing interaction.

To ensure that all observers can communicate the same information about the production process and its interaction with the quantum environment the Kraus partial wave helicity amplitudes (4.32) must transform as two-body helicity $S$-matrix amplitudes~\cite{martin70} under Lorentz transformations. This in turn requires that the spin mixing mechanism is Lorentz invariant.

Let $\Lambda$ be a homogeneous Lorentz transformation from a frame $\Sigma$ to a frame $\Sigma'$. The transformation law of the helicity amplitudes is obtained from
\begin{equation}
U(\Lambda)|p,\mu>=\sum \limits_{\mu'\mu} \mathcal{D}^s_{\mu'\mu}(R)|p',\mu'>
\end{equation}
In (5.7) $p'=\Lambda p$ and $\mathcal{D}^s_{\mu'\mu}(R)$ is the matrix representing Wigner rotation $R(\Lambda,p)=\Lambda^{-1}_{p'} \Lambda \Lambda_p$.
Here $\Lambda_p=R(\theta,\phi)Z_p$ where $Z_p$ is a boost Lorentz transformation along the $z$-axis followed by a rotation through Euler angles $\theta,\phi$. 

With Lorentz symmetry of the $S$-matrix and Kraus operators, the transformation laws for the $S$-matrix and dephasing amplitudes read~\cite{martin70}
\begin{equation}
<p_cp_d;K\mu,\xi|T|p_a p_b;0\nu>=
\end{equation}
\[
\sum \limits_{\mu'\xi'\nu'}(-1)^{\nu'-\nu+\xi'-\xi} 
\mathcal{D}^{K*}_{\mu'\mu}(R_c)
\mathcal{D}^{\frac{1}{2}*}_{\xi'\xi}(R_d) 
<p'_c p'_d; K\mu',\xi'|T|p'_ap'_b;0\nu'>
\mathcal{D}^{\frac{1}{2}}_{\nu'\nu}(R_b)
\]
\begin{equation}
<p_cp_d;J\lambda,\chi|V_\ell|p_cp_d;K\mu,\xi>=
\end{equation}
\[
\sum \limits_{\lambda' \chi'} \sum \limits_{\mu''\xi''}
(-1)^{\xi''-\xi+\chi'-\chi} 
\mathcal{D}^{J*}_{\lambda'\lambda}(R_c)
\mathcal{D}^{\frac{1}{2}*}_{\chi'\chi}(R_d) 
<p'_cp'_d;J\lambda',\chi'|V_\ell|p'_cp'_d;K\mu'',\xi''>
\mathcal{D}^{K}_{\mu''\mu}(R_c)
\mathcal{D}^{\frac{1}{2}}_{\xi''\xi}(R_d)
\]
where we suppressed the isospin labels for the sake of brevity. Combining these results in the expression (4.32) for the Kraus helicity amplitudes and using 
\begin{equation}
\sum \limits_{\mu \xi} 
\mathcal{D}^{K}_{\mu''\mu}(R_c)
\mathcal{D}^{\frac{1}{2}}_{\xi''\xi}(R_d)
\mathcal{D}^{K*}_{\mu'\mu}(R_c)
\mathcal{D}^{\frac{1}{2}*}_{\xi'\xi}(R_d)=\delta_{\mu''\mu'} \delta_{\xi''\xi'}
\end{equation}
we obtain
\begin{equation}
H^J_{\lambda \chi,0\nu}(p_cp_d,\ell)=
\end{equation}
\[
\sum \limits_{\lambda'\chi'} \sum \limits_{\nu'}
(-1)^{\chi'-\chi+\nu'-\nu}
\mathcal{D}^{J*}_{\lambda'\lambda}(R_c)
\mathcal{D}^{\frac{1}{2}*}_{\chi'\chi}(R_d) 
H^J_{\lambda'\chi',0\nu'}(p'_cp'_d,\ell)
\mathcal{D}^{\frac{1}{2}}_{\nu'\nu}(R_b)
\]
This is the required transformation law for the Kraus helicity amplitudes which also shows that the spin mixing mechanism is Lorentz invariant.

The requirement of $P$-parity conservation implies that the Kraus helicity amplitudes satisfy the same parity relations as the corresponding $S$-matrix helisity amplitudes, namely~\cite{svec13a}
\begin{equation}
H^J_{-\lambda -\chi, 0 -\nu}(\ell)=(-1)^{\lambda+\chi+\nu}H^J_{\lambda \chi, 0 \nu}(\ell)
\end{equation}
The relation (4.32) then requires that
\begin{equation}
<p_cp_d;J-\lambda I_J,-\chi|V_\ell|p_cp_d;K-\mu I_K,-\xi>=
\end{equation}
\[
(-1)^{\lambda+\chi-\mu-\xi}
<p_cp_d;J\lambda I_J,\chi|V_\ell|p_cp_d;K\mu I_K,\xi>
\]

Since the dephasing interaction preserves the four-momenta of the all final state particles, the total four-momentum is obviously conserved by Kraus helicity amplitudes. To show they conserve the total angular momentum we consider the $S$-matrix and Kraus amplitudes as functions of the c.m.s. scattering angle $\theta_{c}, \phi_{c}$. Then the partial wave expansion of the final two-body state in $S$-matrix and Kraus amplitudes reads
\begin{eqnarray}
H^K_{\mu \xi,0\nu}(S,\theta_{c} \phi_{c}) & = & 
\sqrt{\frac{4s}{q}}\sum \limits_{J_T,M_T} \sqrt{\frac{2J_T+1}{4\pi}} 
\mathcal{D}^{J_T}_{M_T M'}(\phi_c,\theta_{c},-\phi_{c})
H^K_{\mu \xi,0\nu}(S,J_T M_T)\\
H^J_{\lambda \chi,0\nu}(\ell,\theta_{c} \phi_{c}) & = & 
\sqrt{\frac{4s}{q}}\sum \limits_{J_T,M_T} 
\sqrt{\frac{2J_T+1}{4\pi}} 
\mathcal{D}^{J_T}_{M_T M''}(\phi_c,\theta_{c},-\phi_{c})
H^J_{\lambda \chi,0\nu}(\ell,J_T M_T)
\end{eqnarray}
where $J_T$ and $M_T$ are the total angular mometum and its $z$-axis component, and $M'=\mu-\xi$, $M''=\lambda-\chi$. Assuming $M'=M''$, or
$\lambda-\chi=\mu-\xi$, we can write
\begin{equation}
H^J_{\lambda \chi,0\nu}(\ell,J_T M_T)=\sum \limits_{K\mu,\xi}
<p_cp_d;J\lambda I_J,\chi|V_\ell|p_cp_d;K\mu I_K,\xi>
H^K_{\mu \xi,0\nu}(S,J_T M_T)
\end{equation}
which ensures the conservation of the total angular momentum by the Kraus helicity amplitudes. Note that the equality $M'=M''$ follows from the equation (5.5), i.e. from the Lorentz symmetry of the Kraus operators.

The conservation of the total isospin in Kraus helicity amplitudes follows from the fact that the isospin identities of the final state particles are preserved by the dephasing interaction. Consider $\pi^- p \to \pi^- \pi^+ n$  process. For $K=odd$ isospin state $|I_K>=|1,0>$ while for $K=even$ the isospin state $|I_K>=a|0,0>+b|2,0>$ is a combination of states $|0,0>$ and $|2,0>$. Since the total initial isospin is conserved by the $S$-matrix amplitudes in both cases, then it must be conserved by the dephasing amplitudes with the states $|I_J>=|1,0>$ and $|I_J>=a|0,0>+b|2,0>$ as well. This propagation of total isospin in the spin mechanism ensures isospin conservation by Kraus helicity amplitudes.

Unitarity of the Kraus operators requires
\begin{equation}
\sum \limits_{K\mu,\xi} <p_cp_d;J\lambda I_J,\chi|V_\ell|p_cp_d;K\mu I_K,\xi>
<p_cp_d;J' \lambda' I_{J'},\chi'|V_\ell|p_cp_d;K\mu I_K,\xi>^*=
\delta_{JJ'} \delta_{\lambda \lambda'} \delta_{\chi \chi'}
\end{equation}
The conservation of the total angular momentum constraint (5.5) requires that
$\lambda-\chi=\lambda'-\chi'=\mu-\xi$ for arbitrary $\lambda,\chi,\lambda',\chi'$. For $\lambda-\chi \neq \lambda'-\chi'$ the unitarity relation is a null identity.

\section{Self-consistent form of the spin mixing mechanism.}

\subsection{Kraus transversity amplitudes}

The measurements of $\pi N \to \pi \pi N$ processes on polarized targets are best analyzed in terms of transversity amplitudes with definite $t$-channel naturality. To define Kraus transversity amplitudes we first must define Kraus helicity amplitudes with definite $t$-channel naturality~\cite{svec13a}. The unnatural exchange amplitudes are defined by
\begin{equation}
U^J_{\lambda+,0\pm}(\ell)=\frac{1}{\sqrt{2}} (H^J_{\lambda +,0 \pm}(\ell)+(-1)^\lambda H^J_{-\lambda +,0 \pm}(\ell))
\end{equation}
while the natural exchange amplitudes are given by
\begin{equation}
N^J_{\lambda+,0\pm}(\ell)=\frac{1}{\sqrt{2}} (H^J_{\lambda +,0 \pm}(\ell)-(-1)^\lambda H^J_{-\lambda +,0 \pm}(\ell))
\end{equation}
where $\chi=+$ and $\nu=\pm$ are nucleon helicities. Both amplitudes satisfy the same parity relations (5.12) as the helicity amplitudes. However to express $U^J_{\lambda \chi,0\nu}(\ell)$ and $N^J_{\lambda \chi,0\nu}(\ell)$ in terms of the $S$-matrix amplitudes with definite naturality we need another constraint
\begin{equation}
<p_cp_d;J\lambda I_J,-\chi|V_\ell|p_cp_d;K\mu I_K,-\xi>=(-1)^{\chi-\xi}
<p_cp_d;J\lambda I_J,\chi|V_\ell|p_cp_d;K\mu I_K,\xi>
\end{equation}
which from the $P$-parity relations (5.13) implies
\begin{equation}
<p_cp_d;J-\lambda I_J,\chi|V_\ell|p_cp_d;K-\mu I_K,\xi>=(-1)^{\lambda-\mu}
<p_cp_d;J\lambda I_J,\chi|V_\ell|p_cp_d;K\mu I_K,\xi>
\end{equation}
With the resulting expressions we calculate Kraus unnatural and natural exchange transversity amplitudes defined by~\cite{svec13a}
\begin{eqnarray}
U^J_{\lambda u}(\ell) & = & \frac{1}{\sqrt{2}} (U^J_{\lambda+,0+} (\ell)+iU^J_{\lambda+,0-}(\ell))\\
\nonumber
U^J_{\lambda d}(\ell) & = & \frac{1}{\sqrt{2}} (U^J_{\lambda+,0+}(\ell)-iU^J_{\lambda+,0-}(\ell))
\end{eqnarray}
for unnatural exchange amplitudes and with a similar definition 
\begin{eqnarray}
N^J_{\lambda u}(\ell) & = & \frac{1}{\sqrt{2}} (N^J_{\lambda+,0+} (\ell)+iN^J_{\lambda+,0-}(\ell))\\
\nonumber
N^J_{\lambda d}(\ell) & = & \frac{1}{\sqrt{2}} (N^J_{\lambda+,0+}(\ell)-iN^J_{\lambda+,0-}(\ell))
\end{eqnarray}
for the natural exchange amplitudes. In (6.5) and (6.6) $\tau=u,d$ is the target nucleon transversity. The Kraus transversity amplitudes satisfy parity relations
\begin{eqnarray}
U^J_{-\lambda \tau}(\ell) & = & +(-1)^\lambda U^J_{\lambda \tau}(\ell)\\
\nonumber
N^J_{-\lambda \tau}(\ell) & = & -(-1)^\lambda N^J_{\lambda \tau}(\ell)
\end{eqnarray}
For $\lambda=0$ note that $N^J_{0 +,0 \pm}(\ell)=0$ so that $N^J_{0 \tau}(\ell)=0$.

\subsection{Self-consistency constraints on the spin mixing mechanism}

Starting with the spin mixing mechanism for Kraus helicity amplitudes (4.32) and using the parity relations (5.12), (6.3) and (6.4), the spin mixing mechanism for the Kraus unnatural and natural transversity amplitudes reads
\begin{equation}
U^J_{\lambda \tau}(\ell)=\sum \limits_{K,\mu} <J\lambda|W_\ell|K\mu,\tau>
U^K_{\mu \tau}(S)
\end{equation}
\begin{equation}
N^J_{\lambda \tau}(\ell)=\sum \limits_{K,\mu} <J\lambda|W_\ell|K\mu,\tau>
N^K_{\mu \tau}(S)
\end{equation}
where $U^K_{\lambda \tau}(S)$ and $N^K_{\lambda \tau}(S)$ are the $S$-matrix unnatural and natural exchange transversity amplitudes, respectively, and where
\begin{equation}
 <J\lambda|W_\ell|K\mu,\tau>=
           <p_cp_d;J\lambda,+|V_\ell|p_cp_d;K \mu,+>
\mp i(-1)^\mu <p_cp_d;J\lambda,+|V_\ell|p_cp_d;K-\mu,-> 
\end{equation}
We have omitted isospin labels in (6.10) for the sake of brevity. From the unitarity of the Kraus operators $V_\ell$ follows the unitarity of matrices $<J\lambda|W_\ell|K\mu,\tau>$ in (6.8) 
\begin{equation}
\sum \limits_{K \mu} <J\lambda|W_\ell|K\mu,\tau><J'\lambda'|W_\ell|K\mu,\tau>^*=
\delta_{JJ'} \delta_{\lambda \lambda'}
\end{equation}
The expressions (6.8) and (6.9) satisfy the $P$-parity relations (6.7).

We now impose the conditions $\lambda -\chi=\mu-\xi$ and $K=J-1,J,J+1$ that follow from the Lorentz symmetry of the Kraus operators. The expressions (6.8) and (6.9) then read
\begin{eqnarray}
U^J_{\lambda \tau}(\ell) & = & 
\sum \limits_{K}
<p_cp_d;J\lambda,+|V_\ell|p_cp_d;K \lambda,+>U^K_{\lambda \tau}(S)\\
\nonumber
  &  & \mp i\sum \limits_{K}(-1)^{1-\lambda}<p_cp_d;J\lambda,+|V_\ell|p_cp_d;K\lambda-1,->U^K_{1-\lambda \tau}(S)
\end{eqnarray}
\begin{eqnarray}
N^J_{\lambda \tau}(\ell) & = & 
\sum \limits_{K}
<p_cp_d;J\lambda,+|V_\ell|p_cp_d;K \lambda,+>N^K_{\lambda \tau}(S)\\
\nonumber
  &  & \mp i\sum \limits_{K}(-1)^{1-\lambda}<p_cp_d;J\lambda,+|V_\ell|p_cp_d;K\lambda-1,->N^K_{1-\lambda \tau}(S)
\end{eqnarray}
With a change $\lambda \to -\lambda$ the expressions (6.12) and (6.13) read
\begin{eqnarray}
U^J_{-\lambda \tau}(\ell) & = & 
+(-1)^\lambda \Bigl (\sum \limits_{K}
<p_cp_d;J\lambda,+|V_\ell|p_cp_d;K \lambda,+>U^K_{\lambda \tau}(S)\\
\nonumber
  &  & \mp i\sum \limits_{K}<p_cp_d;J\lambda,+|V_\ell|p_cp_d;K 1+\lambda,-> 
U^K_{1+\lambda \tau}(S) \Bigr )
\end{eqnarray}
\begin{eqnarray}
N^J_{-\lambda \tau}(\ell) & = &  
-(-1)^\lambda \Bigl (\sum \limits_{K}
<p_cp_d;J\lambda,+|V_\ell|p_cp_d;K \lambda,+>N^K_{\lambda \tau}(S)\\
\nonumber
  &  & \pm i\sum \limits_{K}<p_cp_d;J\lambda,+|V_\ell|p_cp_d;K 1+\lambda,-> 
N^K_{1+\lambda \tau}(S) \Bigr )
\end{eqnarray}
These expressions must satisfy the $P$-parity relations (6.7) for any $\lambda$ such that $|\lambda| \leq J$. Equating the second sum in the equations (6.12) and (6.13) with the second sum in (6.14) and (6.15), respectively, we obtain relations that impose spin mixing on the $S$-matrix amplitudes. These constraints are therefore not physical and we must require that
\begin{equation}
<p_cp_d;J\lambda,+|V_\ell|p_cp_d;K\lambda-1,-> =
<p_cp_d;J\lambda,+|V_\ell|p_cp_d;K\lambda+1,-> =0
\end{equation}
Then the spin mechanism for the Kraus transversity amplitudes takes a simple form for any $|\lambda| \leq J$
\begin{equation}
U^J_{\lambda \tau}(\ell)=\sum \limits_{K=J-1,J,J+1}
<p_cp_d;J\lambda I_J,+|V_\ell|p_cp_d;K\lambda I_K,+>
U^K_{\lambda \tau}(S)
\end{equation}
\begin{equation}
N^J_{\lambda \tau}(\ell)=\sum \limits_{K=J-1,J,J+1} 
<p_cp_d;J\lambda I_J,+|V_\ell|p_cp_d;K\lambda I_K,+>
N^K_{\lambda \tau}(S)
\end{equation}
where we have restored the isospin labels and used the constraint (5.6) from the Lorentz symmetry of Kraus operators. The Lorentz conditions (5.5) also require
\begin{equation}
<p_cp_d;J\lambda I_J,+|V_\ell|p_cp_d;K\lambda I_K,->=0
\end{equation}
With the only non-zero dephasing amplitudes given by the spin mixing mechanism (6.17) and (6.18), the spin mixing mechanism for the Kraus helicity amplitudes (4.32) reads
\begin{equation}
H^J_{\lambda \chi, 0 \nu}(\ell)=\sum \limits_{K=J-1,J,J+1}
<p_cp_d;J\lambda I_J,\chi|V_\ell|p_cp_d;K\lambda I_K,\chi>
H^K_{\lambda \chi,0\nu}(S)
\end{equation}
where
\begin{equation}
<p_cp_d;J\lambda I_J,+|V_\ell|p_cp_d;K\lambda I_K,+>=
<p_cp_d;J\lambda I_J,-|V_\ell|p_cp_d;K\lambda I_K,->
\end{equation}
We have arrived at a self-consistent form of the spin mixing mechanism for Kraus amplitudes that arises from their consistency with the Standard Model. The consistency of the Kraus transversity amplitudes with Lorentz symmetry of the Kraus operators results in the breaking of the unitarity of the Kraus operators into a block unitarity with each block characterized by the helicity $\lambda$. The self-consistency constraint requires that the dephasing interaction with the environment conserves both the dipion helicity and the recoil nucleon helicity and transversity.

%\newpage
\section{The measurement of the observed final state $\rho_f(O)$.}

\subsection{Angular intensities and density matrix elements}

Assuming unitary $S$-matrix we used the evolution equation $\rho=S\rho_i S^+$ to develop in Ref~\cite{svec13a} a spin formalism to express the measured final state $\rho_f(p_cp_d;\theta \phi, \vec {P})$ in terms of spin density matrix elements $(R^j_k)^{JJ'}_{\lambda \lambda'}$ which are bilinear combinations of helicity amplitudes $H^J_{\lambda \chi,0\nu}(S)$ or transversity amplitudes $U^J_{\lambda, \tau}(S)$ and $N^J_{\lambda, \tau}(S)$. On the other hand, the data analyses using this formalism lead to the conclusion that $\rho_f(p_cp_d;\theta \phi, \vec{P})$ is given by a non-unitary Kraus representation (4.24). Self-consistency requires that the Kraus representation preserves the experimental form of the angular intensities describing the final state density matrix and the form of the expressions for density matrix elements in terms of transversity amplitudes. In this Section we show that such invariance holds provided that the Kraus partial wave amplitudes satisfy the same parity relations as the $S$-matrix amplitudes due to the conservation of $P$-parity.

The form (4.25) of each term $\rho_f(\theta \phi, \vec{P};\ell\ell)$ in the Kraus representation (4,24) is the same as the form of the $S$-matrix final state in terms of $S$-matrix amplitudes
\begin{equation}
\rho_f(\theta \phi, \vec{P})_{\chi \chi'}=\sum \limits_{\nu \nu'}
S_{\chi,0\nu}(\theta \phi)\rho_b(\vec{P})_{\nu \nu'}
S_{\chi',0\nu'}(\theta \phi)
\end{equation}
We can thus apply the spin formalism develped in Ref.~\cite{svec13a} to each such term separately. Then each term $\rho_f(\theta \phi, \vec{P};\ell\ell)$ in (4.29) has the following form in terms of angular intensities 
\begin{equation}
\rho_f(\theta \phi, \vec{P};\ell\ell) = \frac{1}{2} \bigl (I^0 (\theta \phi, \vec{P};\ell\ell) \sigma^0 + {\vec{I}} (\theta \phi, \vec{P};\ell\ell) \vec{\sigma}\bigr )
\end{equation}
with a decomposition of the intensities $I^j(\theta \phi, \vec{P};\ell\ell)$ in terms of component intensities 
\begin{equation}
I^j(\theta \phi, \vec{P};\ell\ell) 
=I^j_u(\theta \phi;\ell \ell) + P_x I^j_x(\theta \phi;\ell\ell)+ P_y I^j_y(\theta \phi;\ell\ell)+ P_z I^j_z(\theta \phi;\ell\ell)
\end{equation}
The components of intensities $I^j_k(\theta \phi;\ell\ell)$ have angular expansion 
\begin{equation}
I^j_k(\theta \phi;\ell\ell)=\sum \limits_{J \lambda} \sum \limits_{J' \lambda'} (R^j_k(\ell\ell))^{JJ'}_{\lambda \lambda'} Y^J_{\lambda}(\theta \phi) Y^{J'*}_{\lambda'}(\theta \phi)
\end{equation}
where the matrix elements $(R^j_k(\ell\ell))^{JJ'}_{\lambda \lambda'}$ are expressed in terms of Kraus partial wave helicity amplitudes
\begin{equation}
(R^j_k(\ell\ell))^{JJ'}_{\lambda \lambda'}= {1 \over{2}} \sum \limits_{\chi,\chi'} \sum \limits_{\nu \nu'} (\sigma^j)_{\chi' \chi}H^J_{\lambda \chi, 0 \nu}(\ell)(\sigma_k)_{\nu \nu'} H^{J'*}_{\lambda' \chi', 0 \nu'}(\ell)
\end{equation}
As the result of linearity of the Kraus representation, the observed final state density matrix has the same form as in the case of the $S$-matrix
\begin{equation}
\rho_f(\theta \phi, \vec{P}) = {1 \over{2}} \bigl (I^0 (\theta \phi, \vec{P}) \sigma^0 + {\vec{I}} (\theta \phi, \vec{P}) \vec{\sigma}\bigr )
\end{equation}
where
\begin{equation}
I^j_k(\theta \phi)=\sum \limits_{\ell=1}^M  p_{\ell\ell} I^j_k(\theta \phi;\ell\ell)
\end{equation}
To bring $I^j_k(\theta \phi)$ to the experimental $S$-matrix form we need to bring to this form the intensities $I^j_k(\theta \phi;\ell\ell)$. To this end the density matrix elements $(R^j_k(\ell\ell))^{JJ'}_{\lambda \lambda'}$ must satisfy the same parity relations as the corresponding matrix elements in the $S$-matrix theory. This in turn requires that the Kraus partial wave amplitudes satisfy the same parity relations as the corresponding $S$-matrix partial wave amplitudes. With the parity relation (5.12) the components $I^j_k(\theta \phi;\ell\ell)$ then have the desired form~\cite{svec13a} 
\begin{equation}
I^j_k(\theta \phi;\ell\ell) = 
\sum \limits_{J \lambda} \sum \limits _{J' \lambda'} 
(Re R^j_k(\ell \ell))^{JJ'}_{\lambda \lambda'} Re(Y^J_\lambda(\theta\phi)Y^{J'*}_{\lambda'}(\theta \phi))
\end{equation}
for $(k,j)=(u,0),(y,0),(u,2),(y,2),(x,1),(z,1),(x,3),(z,3)$ and the form
\begin{equation}
I^j_k(\theta \phi;\ell\ell) = 
-\sum \limits_{J \lambda} \sum \limits_{J'\lambda'} 
(Im R^j_k(\ell\ell))^{JJ'}_{\lambda \lambda'} 
Im(Y^J_\lambda(\theta \phi)Y^{J'*}_{\lambda'}(\theta \phi))
\end{equation}
for $(x,0),(z,0),(x,2),(z,2),(u,1),(y,1),(u,3),(y,3)$. From (7.7) it follows that the measured intensities $I^j_k(\theta \phi)$ will retain the same form of angular expansions provided that the measured density matrix elements $(R^j_k)^{JJ}_{\lambda \lambda'}$ are redefined as 
\begin{equation}
(R^j_k)^{JJ}_{\lambda \lambda'}=\sum \limits_{\ell=1}^M  p_{\ell\ell}
(R^j_k (\ell\ell))^{JJ'}_{\lambda \lambda'}
\end{equation}
The measured density matrix elements are thus environment-averaged elements of the Kraus density matrices.

\subsection{Effective amplitudes in $\pi N \to \pi \pi N$}

\begin{table}
\caption{Density matrix elements expressed in terms of nucleon transversity amplitudes with definite $t$-channel naturality. The spin indices $JJ'$ which always go with helicities $\lambda \lambda'$ have been omitted in the amplitudes. The coefficients $\eta_\lambda=1$ for $\lambda=0$ and $\eta_\lambda=1/\sqrt{2}$ for $\lambda \neq 0$. Table from Ref.~\cite{svec13a}.}
\begin{tabular}{l|r}
\toprule
$(R^0_u)^{JJ'}_{\lambda \lambda'}$ & $\eta_\lambda \eta_{\lambda'} [U_{\lambda,u}U^*_{\lambda',u}+N_{\lambda,u}N^*_{\lambda',u}
+U_{\lambda,d}U^*_{\lambda',d}+N_{\lambda,d}N^*_{\lambda',d}]$\\
$(R^0_y)^{JJ'}_{\lambda \lambda'}$ & $\eta_\lambda \eta_{\lambda'} [U_{\lambda,u}U^*_{\lambda',u}+N_{\lambda,u}N^*_{\lambda',u}
-U_{\lambda,d}U^*_{\lambda',d}-N_{\lambda,d}N^*_{\lambda',d}]$\\
$(R^0_x)^{JJ'}_{\lambda \lambda'}$ & $-i\eta_\lambda \eta_{\lambda'} [U_{\lambda,u}N^*_{\lambda',d}+N_{\lambda,u}U^*_{\lambda',d}
-U_{\lambda,d}N^*_{\lambda',u}-N_{\lambda,d}U^*_{\lambda',u}]$\\
$(R^0_z)^{JJ'}_{\lambda \lambda'}$ & $\eta_\lambda \eta_{\lambda'} [U_{\lambda,u}N^*_{\lambda',d}+N_{\lambda,u}U^*_{\lambda',d}
+U_{\lambda,d}N^*_{\lambda',u}+N_{\lambda,d}U^*_{\lambda',u}]$\\
\colrule

$(R^2_u)^{JJ'}_{\lambda \lambda'}$ & $-\eta_\lambda \eta_{\lambda'} [U_{\lambda,u}U^*_{\lambda',u}-N_{\lambda,u}N^*_{\lambda',u}
-U_{\lambda,d}U^*_{\lambda',d}+N_{\lambda,d}N^*_{\lambda',d}]$\\
$(R^2_y)^{JJ'}_{\lambda \lambda'}$ & $-\eta_\lambda \eta_{\lambda'} [U_{\lambda,u}U^*_{\lambda',u}-N_{\lambda,u}N^*_{\lambda',u}
+U_{\lambda,d}U^*_{\lambda',d}-N_{\lambda,d}N^*_{\lambda',d}]$\\
$(R^2_x)^{JJ'}_{\lambda \lambda'}$ & $i\eta_\lambda \eta_{\lambda'} [U_{\lambda,u}N^*_{\lambda',d}-N_{\lambda,u}U^*_{\lambda',d}
+U_{\lambda,d}N^*_{\lambda',u}-N_{\lambda,d}U^*_{\lambda',u}]$\\
$(R^2_z)^{JJ'}_{\lambda \lambda'}$ & $-\eta_\lambda \eta_{\lambda'} [U_{\lambda,u}N^*_{\lambda',d}-N_{\lambda,u}U^*_{\lambda',d}
-U_{\lambda,d}N^*_{\lambda',u}+N_{\lambda,d}U^*_{\lambda',u}]$\\
\colrule

$(R^1_u)^{JJ'}_{\lambda \lambda'}$ & $-i\eta_\lambda \eta_{\lambda'} [U_{\lambda,u}N^*_{\lambda',u}-N_{\lambda,u}U^*_{\lambda',u}
-U_{\lambda,d}N^*_{\lambda',d}+N_{\lambda,d}U^*_{\lambda',d}]$\\
$(R^1_y)^{JJ'}_{\lambda \lambda'}$ & $-i\eta_\lambda \eta_{\lambda'} [U_{\lambda,u}N^*_{\lambda',u}-N_{\lambda,u}U^*_{\lambda',u}
+U_{\lambda,d}N^*_{\lambda',d}-N_{\lambda,d}U^*_{\lambda',d}]$\\
$(R^1_x)^{JJ'}_{\lambda \lambda'}$ & $-\eta_\lambda \eta_{\lambda'} [U_{\lambda,u}U^*_{\lambda',d}-N_{\lambda,u}N^*_{\lambda',d}
+U_{\lambda,d}U^*_{\lambda',u}-N_{\lambda,d}N^*_{\lambda',u}]$\\
$(R^1_z)^{JJ'}_{\lambda \lambda'}$ & $-i\eta_\lambda \eta_{\lambda'} [U_{\lambda,u}U^*_{\lambda',d}-N_{\lambda,u}N^*_{\lambda',d}
-U_{\lambda,d}U^*_{\lambda',u}+N_{\lambda,d}N^*_{\lambda',u}]$\\
\colrule

$(R^3_u)^{JJ'}_{\lambda \lambda'}$ & $\eta_\lambda \eta_{\lambda'} [U_{\lambda,u}N^*_{\lambda',u}+N_{\lambda,u}U^*_{\lambda',u}
+U_{\lambda,d}N^*_{\lambda',d}+N_{\lambda,d}U^*_{\lambda',d}]$\\
$(R^3_y)^{JJ'}_{\lambda \lambda'}$ & $\eta_\lambda \eta_{\lambda'} [U_{\lambda,u}N^*_{\lambda',u}+N_{\lambda,u}U^*_{\lambda',u}
-U_{\lambda,d}N^*_{\lambda',d}-N_{\lambda,d}U^*_{\lambda',d}]$\\
$(R^3_x)^{JJ'}_{\lambda \lambda'}$ & $-i\eta_\lambda \eta_{\lambda'} [U_{\lambda,u}U^*_{\lambda',d}+N_{\lambda,u}N^*_{\lambda',d}
-U_{\lambda,d}U^*_{\lambda',u}-N_{\lambda,d}N^*_{\lambda',u}]$\\
$(R^3_z)^{JJ'}_{\lambda \lambda'}$ & $\eta_\lambda \eta_{\lambda'} [U_{\lambda,u}U^*_{\lambda',d}+N_{\lambda,u}N^*_{\lambda',d}
+U_{\lambda,d}U^*_{\lambda',u}+N_{\lambda,d}N^*_{\lambda',u}]$\\

\botrule
\end{tabular}
\label{Table I.}
\end{table}

The Kraus density matrix elements $(R^j_k (\ell\ell))^{JJ'}_{\lambda \lambda'}$ are linear combinations of bilinear terms of Kraus transversity amplitudes which are the same as those for the $S$-matrix elements and which are given in the Table I. The measured density matrix elements $(R^j_k)^{JJ}_{\lambda \lambda'}$ are given by (5.10) in terms of elements of Kraus density matrices. Because this relation is linear the measured elements $(R^j_k)^{JJ}_{\lambda \lambda'}$ will retain the form given in Table I. with moduli and bilinear terms of transversity amplitudes now replaced by environment-averaged moduli and bilinear terms of Kraus amplitudes, respectively. The expressions for the measured bilinear terms for any two transversity amplitudes $A$ and $B$ read
\begin{eqnarray}
<|A|^2>  & = & \sum \limits_{\ell=1}^M p_{\ell\ell}|A(\ell)|^2\\
\nonumber
Re<AB^*> & = &
\sum \limits_{\ell=1}^M  p_{\ell\ell} Re(A(\ell)B^*(\ell))\\
\nonumber
Im<AB^*> & = & \sum \limits_{\ell=1}^M p_{\ell\ell} Im(A(\ell)B^*(\ell))
\end{eqnarray}
where we have omitted the spin labels of the amplitudes for the sake of clarity. Note that the symbol $AB^*$ does not stand for a product of two complex functions $A$ and $B^*$ but communicates which amplitudes $A(\ell)B^*(\ell)$ are being averaged. The averaged bilinear terms can be written in terms of effective transversity amplitudes
\begin{eqnarray}
<|A|^2> & = & |A|^2\\
\nonumber
Re<AB^*> & = & |A||B|\cos\Phi(AB^*)\\
\nonumber
Im<AB^*> & = & |A||B| \sin\Psi(AB^*)
\end{eqnarray}
where $|A|$ and $|B|$ are the moduli of the effective amplitudes and 
$\cos\Phi(AB^*)$ and $ \sin\Psi(AB^*)$ are their correlations. With this form of measured bilinear terms we can no longer associate complex functions $A$ and $B$ with the moduli $|A|$ and $|B|$, respectively, and identify the phases $\Phi(AB^*)$ and $\Psi(AB^*)$ as relative phases of these complex functions. In general the measured correlations violate the necessary trigonometric identity with $\cos^2\Phi(AB^*)+\sin^2\Psi(AB^*)\neq 1$.  Since the phases $\Phi(AB^*)$ are not relative phases of complex functions, they violate the phase condition
\begin{equation}
\Phi(AB^*)=\Phi(AC^*) + \Phi(CB^*)
\end{equation}
and the equivalent cosine condition
\begin{equation}
\cos^2\Phi(AB^*)+\cos^2\Phi(AC^*)+\cos^2\Phi(CB^*)-
2\cos\Phi(AB^*)\cos\Phi(AC^*)\cos\Phi(CB^*)=1
\end{equation}
which hold for any three complex valued functions $A,B,C$.

To better understand the meaning of the measured moduli and correlations we define Kraus vectors in the Hilbert space $H(E)$
\begin{equation}
|A>=\sum \limits_{\ell=1}^M \sqrt{p_{\ell\ell}}A(\ell)|e_\ell>
\end{equation}
and a scalar product
\begin{equation}
<A|B>=\sum \limits_{\ell=1}^M p_{\ell\ell}A(\ell)B^*(\ell)
\end{equation}
Comparing (5.16) with (5.11) and (5.12) we find
\begin{eqnarray}
|A|^2 & = & <A|A>\\
\nonumber
|A||B|\cos\Phi(AB^*) & = & Re<A|B> \\
\nonumber
|A||B|\sin\Psi(AB^*) & = & Im<A|B>
\end{eqnarray}
These relations provide a new physical interpretation of the measured (averaged) bilinear terms: instead of associating the measured effective amplitudes with complex functions we must associate them with complex Kraus vectors (5.15). The measured bilinear terms of effective amplitudes are real or imaginary parts of scalar products of their Kraus vectors. 

It should be noted that the bilinear terms of Kraus amplitudes $A(\ell)B^*(\ell)$ carry the quantum numbers of the state vectors $|e_\ell>$ which define the interacting degrees of freedom of the quantum environment. The bilinear terms of effective amplitudes are mixtures of these quantum numbers.

\subsection{Conservation of probability}

The unitarity of the dephasing amplitudes and the spin mixing mechanism (6.17),(6.18) imply that for each $\ell=1,4$ and $\tau=u,d$
\begin{eqnarray}
\sum \limits_{J=\lambda}^{J_{max}} 
|U^J_{\lambda \tau}(\ell)|^2=
\sum \limits_{K=\lambda}^{J_{max}} 
|U^K_{\lambda \tau}(S)|^2\\
\sum \limits_{J=\lambda}^{J_{max}} 
|N^J_{\lambda \tau}(\ell)|^2=
\sum \limits_{K=\lambda}^{J_{max}}   
|N^K_{\lambda \tau}(S)|^2
\end{eqnarray}
where $J_{max}=J_{max}(m)$ depends on the dipion mass $m$. The observed moduli for transversity amplitudes are given by
\begin{eqnarray}
|U^J_{\lambda \tau}|^2 & = & \sum \limits_{\ell=1}^M p_{\ell\ell}
|U^J_{\lambda \tau}(\ell)|^2\\
\nonumber
|N^J_{\lambda \tau}|^2 & = & \sum \limits_{\ell=1}^M p_{\ell\ell}
|N^J_{\lambda \tau}(\ell)|^2
\end{eqnarray}
It follows from (7.18) and (7.19) that the observed differential cross-section is equal to that given by the $S$-matrix amplitudes
\begin{equation}
\frac{d^2 \sigma}{dmdt} = \sum \limits_{J \lambda,\tau} |U^J_{\lambda \tau}|^2
+|N^J_{\lambda \tau}|^2 = \sum \limits_{J \lambda,\tau} 
|U^J_{\lambda \tau}(S)|^2+|N^J_{\lambda \tau}(S)|^2
\end{equation}
This relation embodies the conservation of probability of a particle reaction in dephasing interaction with the environment. The spin mixing mechanism (6.20) for the helicity amplitudes implies the same conservation of the differential cross-section.

We note that the the observed polarized target assymetry is similarly given by the $S$-matrix amplitudes
\begin{eqnarray}
T \frac{d^2 \sigma}{dmdt} & = & \sum \limits_{J \lambda} 
|U^J_{\lambda u}|^2+|N^J_{\lambda u}|^2-
|U^J_{\lambda d}|^2-|N^J_{\lambda d}|^2\\
\nonumber
  & = & \sum \limits_{J \lambda}
|U^J_{\lambda u}(S)|^2+|N^J_{\lambda u}(S)|^2-
|U^J_{\lambda d}(S)|^2-|N^J_{\lambda d}(S)|^2
\end{eqnarray}
and is conserved by the dephasing interaction. More generally, for two unnatural exchange or two natural exchange transversity amplitudes we have
\begin{equation}
\sum \limits_{J=\lambda}^{J_{max}} A^J_{\lambda \tau}B^{J*}_{\lambda \tau}=
\sum \limits_{K=\lambda}^{J_{max}} A^K_{\lambda \tau}(S)B^{K*}_{\lambda \tau}(S)
\end{equation}
Recall that unitary evolution law of the $S$-matrix imposes relative phases 
$(n(A)-n(B))\pi$ on such transversity amplitudes where $n(A),n(B)$ are integers~\cite{svec13a}. Then the equations (7.23) imply
\begin{equation}
\sum \limits_{J=\lambda}^{J_{max}} |A^J_{\lambda \tau}||B^J_{\lambda \tau}|
\sin \Psi(A^J_{\lambda \tau}B^J_{\lambda \tau})=0
\end{equation}

\subsection{Determination of the amplitudes in measurements on polarized targets}

The CERN-Cracow-Munich (CCM) measurements of $\pi^- p \to \pi^- \pi^+ n$ at 17.2 GeV/c~\cite{becker79a,becker79b,chabaud83,rybicki85} measured moments $t^L_M$ (unpolarized target), $p^L_M$  (polarization component perpendicular to the scattering plane) and $r^L_M$ (polarization component in the scattering plane).
These moments are linear combinations of spin density matrix elements $(\rho^0_u)^{JJ'}_{\lambda \lambda'},(\rho^0_y)^{JJ'}_{\lambda \lambda'}$ and 
$(\rho^0_x)^{JJ'}_{\lambda \lambda'}$, respectively, corresponding to different spins $J,J'$. The ITEP measurements of $\pi^- p \to \pi^- \pi^+ n$ at 1.78 GeV/c ~\cite{alekseev99} and the CERN-Saclay measurements of $\pi^+ p \to \pi^+ \pi^- p$ at 5.98 and 11.85 GeV/c~\cite{lesquen85} measured directly the spin density matrix elements $(\rho^0_u)^{JJ'}_{\lambda \lambda'},(\rho^0_y)^{JJ'}_{\lambda \lambda'}$ and $(\rho^0_x)^{JJ'}_{\lambda \lambda'}$ for the $S$- and $P$-wave subsystem.

To determine the observed production amplitudes CCM used a $\chi^2$ minimization method~\cite{becker79a,becker79b,chabaud83,rybicki85,kaminski97}. The moments are expressed in terms of the transversity amplitudes and these theoretical formulas are fitted to the measured momements. All $\chi^2$ fits found two solutions  $A_u(i),A_d(i),i=1,2$ for the $S$- and $P$-wave amplitudes $A=S,P$ below 980 MeV ($K\bar{K}$ threshold) and a single solution for $S$-,$P-$ and $D$-waves above 980 MeV. The cosines of the relative phases are treated as free parameters and are not constrained by the cosine conditions (3.11) in the minimization process.

The combinations $(\rho^0_u)^{JJ'}_{\lambda \lambda'} \pm (\rho^0_y)^{JJ'}_{\lambda \lambda'}$ of the density matrix elements of the $S$- and $P$-wave subsystem form two separate systems of equations for the amplitudes with transversity "up" (sign +) and "down" (sign -). Each system is analytically solvable provided the cosines of the relative phases in each system satisfy the cosine conditions (3.11) which leads to a cubic equation for each $|L_\tau|^2$.
The two real physical solutions of each cubic equation allow to determine two solutions $A_u(i),A_d(j),i,j=1,2$ for the $S$- and $P$-wave amplitudes $A=S,P$ below 980 MeV. 

The analytical method has been used to determine the initial values in the $\chi^2$ minimization analyses~\cite{becker79a,becker79b,chabaud83,rybicki85,kaminski97}. It has been fully employed in the ITEP analysis~\cite{alekseev99} and in the Monte Carlo analyses of the CCM data on $\pi^- p \to \pi^- \pi^+ n$ at 17.2 GeV/c and the CERN-Saclay data on $\pi^+ p \to \pi^+ \pi^- p$ at 5.98 and 11.85 GeV/c~\cite{svec96,svec97a,svec12a}. 

In Monte Carlo analyses the amplitudes are calculated for each sampling of the error volume. Unphysical solutions are rejected and the physical solutions are used to determine the average values of the amplitudes and their errors. The resulting amplitudes are genuine complex valued functions. 

Kraus amplitudes are complex valued functions. This suggests that the $S$- and $P$-wave analytical amplitudes may be related to the Kraus amplitudes. Since the Monte Carlo amplitude analysis selects data that satisfy the cosine conditions, this data must describe a decoherence free subspace. In this case there is only one set of $S$- and $P$-wave Kraus amplitudes equal to one of the two analytical solutions. In our new analysis of the CERN data using spin mixing mechanism~\cite{svec14a} we show that spin mixing mechanism selects the Solution 2 of the analytical as well as the $\chi^2$ methods.

There are no exact analytical solutions of the $S$-, $P$- and $D$-wave system. We thus expect a single effective solution for the $S$-,$P$-and $D$-wave system with some cosines of relative phases violating the cosine conditions. In the Section IX. we present evidence from the analysis of Rybicki and Sakrejda~\cite{rybicki85} that the $D$-waves indeed violate the cosine condition.

%\newpage
\section{The prediction of the $\rho^0(770)-f_0(980)$ mixing in $\pi^- p \to \pi^- \pi^+ n$.}

\subsection{Spin mixing mechanism in  $\pi^- p \to \pi^- \pi^+ n$ below 1400 MeV}

Below 1400 MeV the process $\pi^- p \to \pi^- \pi^+ n$ is described by $S$-, $P$- and $D$-wave transversity amplitudes. In the following we shall use a new notation for these amplitudes. The amplitudes $S_\tau$ and $L_\tau$ are the $S$-wave and $P$-wave unnatural exchange amplitudes with helicity $\lambda=0$. The amplitudes $U_\tau$ and $N_\tau$ are the $P$-wave unnatural and natural exchange amplitudes with helicity $\lambda=1$. There are three unnatural exchamge $D$-wave amplitudes $D^0_\tau,D^U_\tau,D^{2U}_\tau$ corresponding to the helicities $\lambda=0,1,2$ and two natural exchange $D$-wave amplitudes $D^N_\tau,D^{2N}_\tau$ corresponding to helicities $\lambda=1,2$.

With the notation $V^\lambda_{JK}(\ell)=<p_cp_d;J\lambda I_J,+|V_\ell|p_cp_d;K\lambda I_K,+>$ in the spin mixing mechanism (6.17) and (6.18) the expressions for the $S$-,$P$- and $D$-wave Kraus amplitudes in terms of $S$-matrix amplitudes read as follows. The three helicity zero amplitudes $S_\tau,L_\tau,D^0_\tau$ form a dephasing triplet 
\begin{eqnarray}
D^0_\tau(\ell) & = & V^0_{20}(\ell)S_\tau(S)+V^0_{21}(\ell)L_\tau(S)+V^0_{22}(\ell)D^0_\tau(S)\\
\nonumber
L_\tau(\ell) & = & V^0_{10}(\ell)S_\tau(S)+V^0_{11}(\ell)L_\tau(S)+V^0_{12}(\ell)D^0_\tau(S)\\
\nonumber
S_\tau(\ell) & = & V^0_{00}(\ell)S_\tau(S)+V^0_{01}(\ell)L_\tau(S)+V^0_{02}(\ell)D^0_\tau(S)
\nonumber
\end{eqnarray}
The amplitudes $U_\tau,D^U_\tau$ and $N_\tau,D^N_\tau$ form two dephasing doublets. 
\begin{eqnarray}
D^U_\tau(\ell) & = & V^1_{21}(\ell)U_\tau(S)+V^1_{22}(\ell)D^U_\tau(S)\\
\nonumber
U_\tau(\ell) & = & V^1_{11}(\ell)U_\tau(S)+V^1_{12}(\ell)D^U_\tau(S)
\nonumber
\end{eqnarray}
\begin{eqnarray}
D^N_\tau(\ell) & = & V^1_{21}(\ell)N_\tau(S)+V^1_{22}(\ell)D^N_\tau(S)\\
\nonumber
N_\tau(\ell) & = & V^1_{11}(\ell)N_\tau(S)+V^1_{12}(\ell)D^N_\tau(S)
\nonumber
\end{eqnarray}
The amplitudes $D^{2U}$ and $D^{2N}$ form two dephasing singlets
\begin{eqnarray}
D^{2U}_\tau(\ell) & = & V^2_{22}(\ell)D^{2U}_\tau(S)\\
\nonumber
D^{2N}_\tau(\ell) & = & V^2_{22}(\ell)D^{2N}_\tau(S)
\nonumber
\end{eqnarray}
The matrices in (8.1)-(8.4) are unitary matrices $U(3),U(2)$ and $U(1)$.

Lorentz symmetry of Kraus operators imposes a constraint $K=J-1,J,J+1$. This means that $V^0_{20}=V^0_{02}=0$ in (8.1). The unitarity of this matrix then implies $V^0_{21}(\ell)=V^0_{12}(\ell)=0$ and (8.1) takes the form
\begin{eqnarray}
D^0_\tau(\ell) & = & V^0_{22}(\ell)D^{0}_\tau(S)\\
\nonumber
L_\tau(\ell) & = & V^0_{10}(\ell)S_\tau(S)+V^0_{11}(\ell)L_\tau(S)\\
\nonumber
S_\tau(\ell) & = & V^0_{00}(\ell)S_\tau(S)+V^0_{01}(\ell)L_\tau(S)
\end{eqnarray}
$D^0_\tau(\ell)$ is thus a dephasing singlet and the amplitudes $L_\tau(\ell)$ and $S_\tau(\ell)$ form a dephasing doublet. The amplitude $D^0_\tau(\ell)$ decouples from the $S$- and $P$- subsystem with the result that there can be no spin mixing of $D^0_\tau(S)$ and $L_\tau(S)$ amplitudes.  The matrix in (8.6) is a $U(2)$ matrix. 

In the mass interval below 980 MeV at low $t$ the $S$- and $P$-waves dominate and $D$-waves are neglected. There is no evidence for the presence of $\rho^0(770)$ in the measured $D^0$ at high $t$ as predicted by the equations (8.5). The observed amplitudes $U_\tau$ and $N_\tau$ resonate at $\rho^0(770)$ mass and at high $t$ $N_\tau$ are the largest amplitudes. There is no evidence for the presence of $\rho^0(770)$ in the amplitudes $D^U$ and $D^N$ at high $t$. This implies $V^1_{12}(\ell)= V^1_{21}(\ell)=0$. Then (8.2) reads
\begin{eqnarray}
D^U_\tau(\ell) & = & V^1_{22}(\ell)D^U_\tau(S)=\exp{i\eta(\ell)}D^U_\tau(S)\\
\nonumber
U_\tau(\ell) & = & V^1_{11}(\ell)U_\tau(S)=\exp{i\delta(\ell)}U_\tau(S)
\nonumber
\end{eqnarray}
In this mass range there is a consistent body of evidence for $\rho^0(770)-f_0(980)$ mixing in the amplitudes $S_d$ and $L_d$ (for a review see Ref.~\cite{svec12d}).

\subsection{Phases of the $S$-matrix amplitudes}

To study the observable effects of the spin mixing mechanism we need to introduce the phases of the $S$-matrix transversity amplitudes. In this subsection we omit the label $S$ of the $S$-matrix transversity amplitudes.

In the Part I. of this work~\cite{svec13a} we have shown that the unitary evolution law imposes a constraint on the relative phases of the $S$-matrix amplitudes
\begin{equation}
\Phi(U^J_{\lambda \tau})-\Phi(N^{J'}_{\lambda' -\tau})=0,\pm \pi, \pm2\pi
\end{equation}
for all $J,J'$ and corresponding $\lambda,\lambda'$. It follows from the unitary conditions (8.7) that all unnatural exchange amplitudes $U^J_{\lambda \tau}$ as well as all natural exchange amplitudes $N^J_{\lambda \tau}$ share the same absolute phase up to an integer multiple of $\pi$
\begin{eqnarray}
\Phi(U^J_{\lambda \tau}) & = & \Phi(S_\tau)+\pi n(U^J_{\lambda \tau})\\
\nonumber
\Phi(N^J_{\lambda \tau}) & = & \Phi(N_\tau)+\pi n(N^J_{\lambda \tau})
\end{eqnarray}
where $n(U^J_{\lambda \tau})$ and $n(N^J_{\lambda \tau})$ are integers. The phases $\Phi(S_\tau)$ and $\Phi(N_\tau)$ depend on $s,t,m$. 

The unitary phases (8.8) imply that if there is a resonance at the mass $m_R$ in the amplitudes $U^{J_R}_{\lambda \tau}$  ($N^{J_R}_{\lambda \tau}$) with $J=J_R$, then all contributing amplitudes $U^{J}_{\lambda \tau}$ 
($N^{J}_{\lambda \tau}$)  with $J \neq J_R$ will have the same resonant phase near $m_R$. This does not imply that any of the amplitudes with $J \neq J_R$ must resonate and show a resonant peak or a dip at $m_R$ since their non-resonant moduli are mathematically allowed to have resonant phases. Although the unitary phases (8.8) appear to allow such resonance mixing, it is excluded by the the symmetries of the $S$-matrix and Standard Model. Indeed, the production as well as the decay of a resonance is described entirely by the resonant $S$-matrix amplitudes $A^{J_R}_{\lambda \tau}(S)$ while the mixing of resonances in the Kraus amplitudes $A^J_{\lambda \tau}(\ell)$ arises entirely from the dephasing amplitudes.

With (8.8) the $S$-matrix transversity amplitudes then have the form
\begin{eqnarray}
U^J_{\lambda \tau} & = & e^{i\Phi(S_\tau)} e^{i\pi n(U^J_{\lambda \tau})} 
|U^J_{\lambda \tau}|\\
\nonumber
N^J_{\lambda \tau} & = & e^{i\Phi(N_\tau)} e^{i\pi n(N^J_{\lambda \tau})}
|N^J_{\lambda \tau}|
\end{eqnarray}
A self-consistent set of relative phases is presented in the Table II. in the Part I. of this work~\cite{svec13a}. For the $S$- and $P$-wave subsystem below 980 MeV the relative phases read
\begin{subequations}
\begin{eqnarray}
\Phi(L^0_\tau)-\Phi(S^0_\tau) & = 0   \\
\Phi(L^0_\tau)-\Phi(U^0_\tau) & = +\pi \\
\Phi(U^0_\tau)-\Phi(S^0_\tau) & = -\pi
\end{eqnarray}
\end{subequations}
For masses $m \gtrsim 980$ MeV there is a change of sign in the measured interference terms of the amplitudes prompting a change of relative phases for $\tau=d$~\cite{svec13a}
\begin{subequations}
\begin{eqnarray}
\Phi(L^0_d)-\Phi(S^0_d) & =+\pi \\
\Phi(L^0_d)-\Phi(U^0_d) & =+\pi \\
\Phi(U^0_d)-\Phi(S^0_d) & =0
\end{eqnarray}
\end{subequations}
In (8.10) and (8.11) we have introduced the superscript $0$ to label the $S$- and $P$-wave $S$-matrix amplitudes to be used in the next subsection. It follows that for $m < 980$ MeV
\begin{subequations}
\begin{eqnarray}
S^0_\tau & = & +e^{i\Phi(S_\tau)} |S^0_\tau|\\
L^0_\tau & = & +e^{i\Phi(S_\tau)} |L^0_\tau|\\
U^0_\tau & = & -e^{i\Phi(S_\tau)} |U^0_\tau|
\end{eqnarray}
\end{subequations}
while for the masses above 980 MeV the signs of $L^0_d $ and $U^0_d$ are reversed.

\subsection{Predictions arising from the $\rho^0(770)-f_0(980)$ mixing}

The spin mixing matrix $V^0_{JK}$ in (8.6) is a unitary $U(2)$ matrix. The most general $U(2)$ matrix has the form~\cite{seiden05}
\begin{eqnarray}
V^0_{JK} & = & \left(
\begin{array}{cc}
+e^{i\phi_1} \cos \theta & e^{i(\phi_1+\phi_2)} \sin \theta \\
-e^{i\phi_2} \sin \theta & e^{i(\phi_2+\phi_3)} \cos \theta
\end{array} \right)
\end{eqnarray}
With $V^1_{11}=e^{i\delta}$ the spin mixing mechanism for $S$- and $P$-wave Kraus amplitudes $\ell=1,M$ reads
\begin{subequations}
\begin{eqnarray}
L_\tau (\ell) & = & e^{i\phi_1(\ell)} \bigl ( +\cos \theta(\ell) S^0_\tau + e^{i\phi_2(\ell)} \sin\theta(\ell) L^0_\tau \bigr )\\
S_\tau (\ell) & = & e^{i\phi_2(\ell)} \bigl ( -\sin \theta(\ell) S^0_\tau + e^{i\phi_3(\ell)} \cos\theta(\ell) L^0_\tau \bigr )\\
U_\tau (\ell)& = & e^{i\delta(\ell)} U^0_\tau\\
N_\tau (\ell)& = & e^{i\delta(\ell)} N^0_\tau
\end{eqnarray}
\end{subequations}

In the following we omit the label $\ell$. First we consider the $\rho^0(770)$ mass region and $m < 980$ MeV. With the phases (8.12) the equations (8.14) read
\begin{subequations}
\begin{eqnarray}
L_\tau & = & +e^{i\Phi(S_\tau)} e^{i\phi_1} \bigl ( +\cos \theta |S^0_\tau| + e^{i\phi_2} \sin\theta |L^0_\tau| \bigr )\\
S_\tau & = & +e^{i\Phi(S_\tau)} e^{i\phi_2} \bigl ( -\sin \theta |S^0_\tau| + e^{i\phi_3} \cos\theta |L^0_\tau| \bigr )\\
U_\tau & = & -e^{i\Phi(S_\tau)} e^{i\delta(U)} |U^0_\tau|
\end{eqnarray}
\end{subequations}
The moduli have a form
\begin{eqnarray}
|L_\tau|^2 & = & \sin^2 \theta |L^0_\tau|^2 \bigl( 1+2\cot \theta \cos \phi_2 X_\tau +\cot^2 \theta X^2_\tau \bigr)\\
\nonumber
|S_\tau|^2 & = & \cos^2 \theta |L^0_\tau|^2 \bigl( 1-2\tan \theta \cos \phi_3 X_\tau +\tan^2 \theta X^2_\tau \bigr)\\
\nonumber
|U_\tau|^2 & = & |U^0_\tau|^2
\end{eqnarray}
where $X_\tau=|S^0_\tau|/|L^0_\tau|$. To fix the phases $\phi_i,i=1,3$ we require that the cosine $\cos \Phi_\tau(LS)$ of the relative phase $\Phi_\tau(LS)=\Phi(L_\tau)-\Phi(S_\tau)$ is near its unitary value $\cos \Phi^0_\tau(LS)=+1$ given by (8.10a)
\begin{equation}
\cos \Phi_\tau(LS) = \frac {Re(L_\tau S^*_\tau)}{|L_\tau||S_\tau|} = 1 -\epsilon
\end{equation}
From (8.15) and (8.16) we find
\begin{eqnarray}
\cos \Phi_\tau(LS) & = & \frac {\cos(\phi_1-\phi_3)\sin\theta \cos\theta |L^0_\tau|^2 (1+...)}{|\sin\theta||\cos\theta||L^0_\tau|^2(1+...)}\\
\nonumber
  & = & \cos(\phi_1-\phi_3)(1-\epsilon)
\end{eqnarray}
where we have assumed $\sin\theta>0,\cos\theta>0$. The consistency with the unitary value requires that $\phi_1=\phi_3=\phi$. The unitarity of the resulting $V^0_{JK}$ requires $\phi_2=\phi$. Similarly we require that 
\begin{equation}
\cos\Phi_\tau(LU)=\frac{Re(L_\tau U^*_\tau)}{|L_\tau||U_\tau|}=-1+\epsilon
\end{equation}
is near the unitary value $\cos\Phi^0_\tau(LU)=-1$. This consistency implies that $\delta=2\phi$. The final form of the spin mixing mechanism below 980 MeV then reads
\begin{eqnarray}
L_\tau & = & +e^{i\Phi(S_\tau)} e^{i\phi} \bigl ( +\cos \theta |S^0_\tau| + e^{i\phi} \sin\theta |L^0_\tau| \bigr )\\
\nonumber
S_\tau & = & +e^{i\Phi(S_\tau)} e^{i\phi} \bigl ( -\sin \theta |S^0_\tau| + e^{i\phi} \cos\theta |L^0_\tau| \bigr )\\
\nonumber
U_\tau & = & -e^{i\Phi(S_\tau)} e^{i2\phi} |U^0_\tau|
\end{eqnarray}
With $\phi_2=\phi_3=\phi$ the equations (8.16) take the form
\begin{eqnarray}
|L_\tau|^2 & = & \cos^2 \theta |S^0_\tau|^2 +\sin^2 \theta |L^0_\tau|^2 +\sin 2\theta \cos\phi|S^0_\tau||L^0_\tau|\\
\nonumber
|S_\tau|^2 & = & \sin^2 \theta |S^0_\tau|^2 +\cos^2 \theta |L^0_\tau|^2 -\sin 2\theta \cos\phi|S^0_\tau||L^0_\tau|\\
\nonumber
|U_\tau|^2 & = & |U^0_\tau|^2
\end{eqnarray}
and predict the presence of $\rho^0(770)$ resonance in the amplitudes $|S_\tau|^2$. 

We now consider the $f_0(980)$ mass region $m \gtrsim 980$ MeV. We shall focus on amplitudes with transversity $\tau=d$. Since there is no change in the amplitudes with $\tau=u$ we assume we still have $\phi_1=\phi_2=\phi_3=\phi$ and $\delta=2 \phi$. With the reversed signs of $L_d$ and $U_d$ in (8.12) the final form of the spin mixing mechanism (8.14) for $\tau=d$ reads
%\begin{subequations}
\begin{eqnarray}
L_d & = & -e^{i\Phi(S_d)} e^{i\phi} \bigl ( -\cos \theta |S^0_d| + e^{i\phi} \sin\theta |L^0_d| \bigr )\\
\nonumber
S_d & = & -e^{i\Phi(S_d)} e^{i\phi} \bigl ( +\sin \theta |S^0_d| + e^{i\phi} \cos\theta |L^0_d| \bigr )\\
\nonumber
U_d & = & +e^{i\Phi(S_d)} e^{i2 \phi} |U^0_d|
\end{eqnarray}
%\end{subequations}
The moduli for $\tau=d$ have the form
\begin{eqnarray}
|L_d|^2 & = & \cos^2 \theta |S^0_d|^2 +\sin^2 \theta |L^0_d|^2 -\sin 2\theta \cos\phi |S^0_d||L^0_d|\\
\nonumber
|S_d|^2 & = & \sin^2 \theta |S^0_d|^2 +\cos^2 \theta |L^0_d|^2 +\sin 2\theta \cos\phi |S^0_d||L^0_d|\\
\nonumber
|U_d|^2 & = & |U^0_d|^2
\end{eqnarray}
For $\tau=u$ the moduli retain the form (8.21). The equations (8.21) and (8.23) both predict the presence of $f_0(980)$ resonance in the amplitudes $|L_\tau|^2$.

Expressions for the relative phases can be solved exactly to read
\begin{eqnarray}
\tan \Phi_\tau(LS)=\frac{Im(L_\tau S^*_\tau)}{Re(L_\tau S_\tau^*)} & = &
\frac{-2 \sin \phi X_\tau}{(1+2\cot2\theta \cos\phi X_\tau -X_\tau^2)
\sin 2\theta }\\
\tan \Phi_\tau(LU)=\frac{Im(L_\tau U^*_\tau)}{Re(L_\tau U_\tau^*)} & = &
\frac{-\sin \phi \cot \theta X_\tau}{1+\cot \theta \cos \phi X_\tau}\\
\tan \Phi_\tau(US)=\frac{Im(U_\tau S^*_\tau)}{Re(U_\tau S_\tau^*)} & = &
\frac{-\sin \phi \tan \theta X_\tau}{1-\tan \theta \cos \phi X_\tau}
\end{eqnarray}
where $\Phi_\tau(AB)=\Phi (A_\tau)-\Phi (B_\tau)$. From the ratio
\begin{equation}
Y_\tau=\frac{\tan \Phi_\tau(LU)}{\tan \Phi_\tau(US)}=\cot^2 \theta \frac{1-\tan \theta \cos \phi X_\tau}{1+\cot \theta \cos \phi X_\tau}
\end{equation}
we obtain a quadratic equation for the cotangent $\cot \theta$ which has two solutions 
\begin{equation}
\cot \theta_{1,2} = \frac{1}{2} \Bigl( \cos \phi (1+Y_\tau)X_\tau \pm 
\sqrt{\bigl(\cos \phi (1+Y_\tau)X_\tau \bigr)^2 + 4Y_\tau } \Bigr)
\end{equation}
Only $\cot \theta >0$ is a physical solution for the assumed $\cos \theta >0, \sin \theta>0$.

The spin mixing parameters $\theta (s,m)$ and $\phi (s,m)$ depend on the c.m.s. energy $s$ and the invariant dipion mass $m$ but do not depend on the transversity $\tau$ and momentum transfer $t$. For the expression on the r.h.s. of equation (8.28) to be independent of $\tau$ and $t$ for all $s,m,t$ we need $Y_\tau$ and $X_\tau$ to be independent of $\tau$ and $t$. This independence for $X_\tau$ requires proportionality relations for the $S$-matrix amplitudes
\begin{eqnarray}
|S^0_u| & = & K |S^0_d|\\
\nonumber
|L^0_u| & = & K |L^0_d|
\end{eqnarray}
and a factorization of their mass and momentum transfer dependence
\begin{eqnarray}
|S^0_\tau(s,m,t)|=F^0_\tau(s,m)Q_\tau(s,m,t)\\
\nonumber
|L^0_\tau(s,m,t)|=G^0_\tau(s,m)Q_\tau(s,m,t)
\end{eqnarray}
The spin mixing mechanism (8.20) and (8.22) then implies the same relations for the spin mixing amplitudes $S_\tau(s,m,t)$ and $L_\tau(s,m,t)$ with the same $Q_\tau (s,m,t)$. These relations ensure the required independence of $Y_\tau$. The relations (8.29) and (8.30) do not apply to the helicity $\lambda=1$ amplitudes $|U^0_\tau|$ and $|N^0_\tau|$.

The independence of $X_\tau$ on $\tau$ and $t$ means that the relative phases do not depend on the transversity and, at a fixed $s$ and $m$, on $t$. The spin mixing mechanism thus predicts that the solutions for the cosines of relative phases in amplitude analyses should be the same for both transversities and independent of $t$ at the same energy.

\subsection{A qualitative comparison of the predictions with the experimental data}

There are $M=\dim H(E)$ Kraus amplitudes $S_\tau(\ell)$ and $L_\tau(\ell)$. The spin mixing parameters $\theta(\ell)$ and $\phi(\ell)$ are functions of energy $s$ and dipion mass $m$ but do not depend on the momentum transfer $t$ and the transversity $\tau$. We can identify the complex valued functions $S_\tau(\ell)$ and $L_\tau(\ell)$ with the complex valued analytical amplitudes obtained from the experimental data. 

There are two solutions for the moduli of the $S$- and $P$-wave amplitudes
and the cosines of their relative phases which implies $M=2$. It is also possible to cross combine the Solution 1 (2) for the $\tau=u$ amplitudes with the Solution 2 (1) for the $\tau=d$ amplitudes. Such four combinations of the solutions $11$, $12$, $21$ and $22$ imply $M=4$. Since the cosines do not determine the signs of the relative phases, there is a four-fold sign ambiguity $++,+-,-+,--$ in the relative signs of the relative phases in each solution in both $M=2$ and $M=4$ cases. The amplitudes of the solutions $--$ and $-+$ are simply complex conjugate amplitudes of the solutions $++$ and $+-$, respectively. The cross solutions $12$ and $21$ are excluded by the spin mixing mechanism which requires $|S_u(i)|^2=K^2|S_d(i)|^2$ and $|L_u(i)|^2=K^2|L_d(i)|^2$ for each Solution $i=1,2$. In the following we shall focus on the $M=2$ case with Solutions 1 and 2 corresponding to combinations $11$ and $22$. 

At this point we do not know the $S$-matrix amplitudes $S_\tau^0$ and $L_\tau^0$, i.e. the ratio $X_\tau$, and the spin mixing parameters $\theta$ and $\phi$ so we cannot make quantitative predictions for the moduli (8.23) and the cosines (8.24)-(8.26). However we can still make qualitative comparisons with the data and draw from them important conclusions. 

\subsubsection{Predictions for the moduli}

Figure 1 shows the $S$-wave moduli from the recent Monte Carlo analysis of the CERN-Cracow-Munich data (CCM) at low momentum transfers $0.005 \leq |t| \leq 0.20$ (GeV/c)$^2$~\cite{svec12a}. These results are nearly identical to the published results of the $\chi^2$ fit analysis of the same data by the CCM group~\cite{kaminski97} shown in the Figure 2. Both analyses clearly indicate the presence of a $\rho^0(770)$ peak in both solutions for the amplitude $|S_d|^2$. The Solution 2 for $|S_d|^2$ exhibits a pronounced structure at ~930 MeV  absent in the Solution 1.

The results of the recent Monte Carlo analysis~\cite{svec12a} are similar to the previously published Monte Carlo amplitude analyses of the CCM data at smaller Monte Carlo resolution~\cite{svec96,svec97a}. The recent work includes analysis on unpolarized target and the determination of helicity amplitudes which both confirm the presence of the $\rho^0(770)$ in the $S$-wave. The 1997 $\chi^2$ fit analysis~\cite{kaminski97} is in agreement with the 1979 $\chi^2$ analysis using CCM data from a different run~\cite{becker79a}. Independent evidence for $\rho^0(770)$ in the $S$-wave comes from the analysis of ITEP data on $\pi^- p \to \pi^- \pi^+ n$ at 1.78 GeV/c at low $t$~\cite{alekseev99}, and from two analyses at high $t$ of CCM data~\cite{rybicki85} and CERN-Saclay data 
on $\pi^+ n \to \pi^+ \pi^- p$ at 5.98 and 11.85 GeV/c~\cite{svec96,svec97a,svec12a}. For a full review with figures of all these results for the moduli and intensities see Ref.~\cite{svec12d}. This survey shows a remarkable consistency of the evidence for a rho-like state in the $S$-wave in $\pi^- \pi^+$ production channel.

Spin mixing mechanism explains the appearance of the $\rho^0(770)$ peak in the measured amplitudes $|S_d|^2$ as due to the presence of a large resonating $S$-matrix amplitude $|L_d^0|^2$ in the expression for the $S$-wave modulus (8.23) for sufficiently large spin mixing parameter $\cos^2 \theta$. In CCM data this spin mixing appears suppressed in the amplitudes $|S_u|^2$ compared to $|S_d|^2$. 

The Monte Carlo analysis of CCM data~\cite{svec12a} shows a dip in the amplitude $|L_d|^2$ near the $f_0(980)$ mass in both solutions while there is a sudden drop in $|S_d|^2$ at this mass. These effects are due to the mixing of the decreasing $S$-matrix amplitude $L^0_d$ and increasing $S$-matrix amplitude $S^0_d$ above the $K\bar{K}$ threshold at 980 MeV, indicating $\rho^0(770)-f_0(980)$ mixing in the $P$-wave amplitude.

\begin{figure} [htp]
\includegraphics[width=12cm,height=10.5cm]{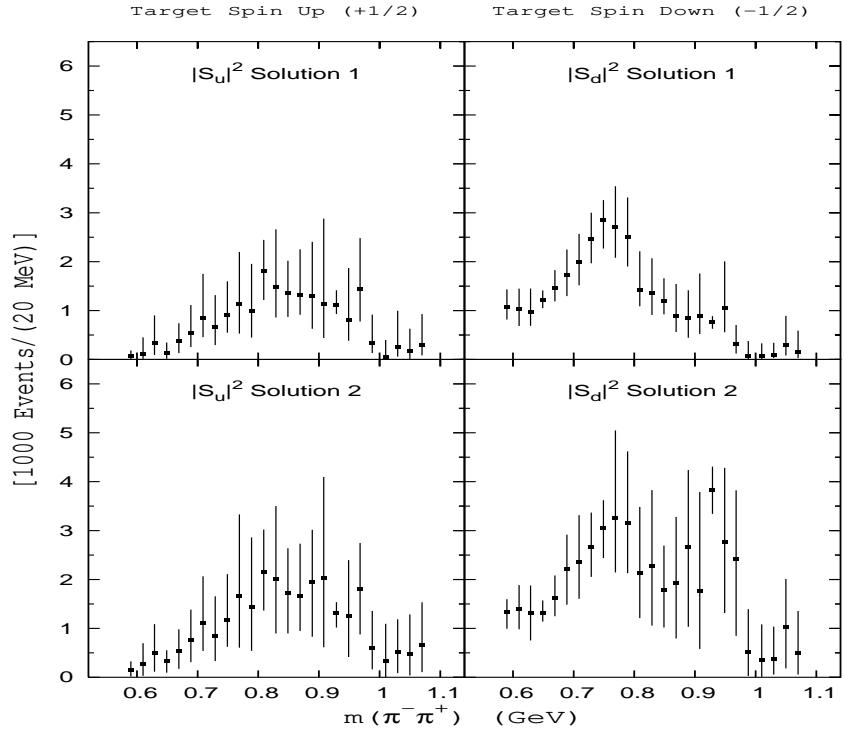}
\caption{Moduli of $S$-wave amplitudes $|S_u|^2$ and $|S_d|^2$ from the Monte Carlo analysis~\cite{svec12a}.}
\label{Figure 1}
\end{figure}

\begin{figure} [hp]
\includegraphics[width=12cm,height=10.5cm]{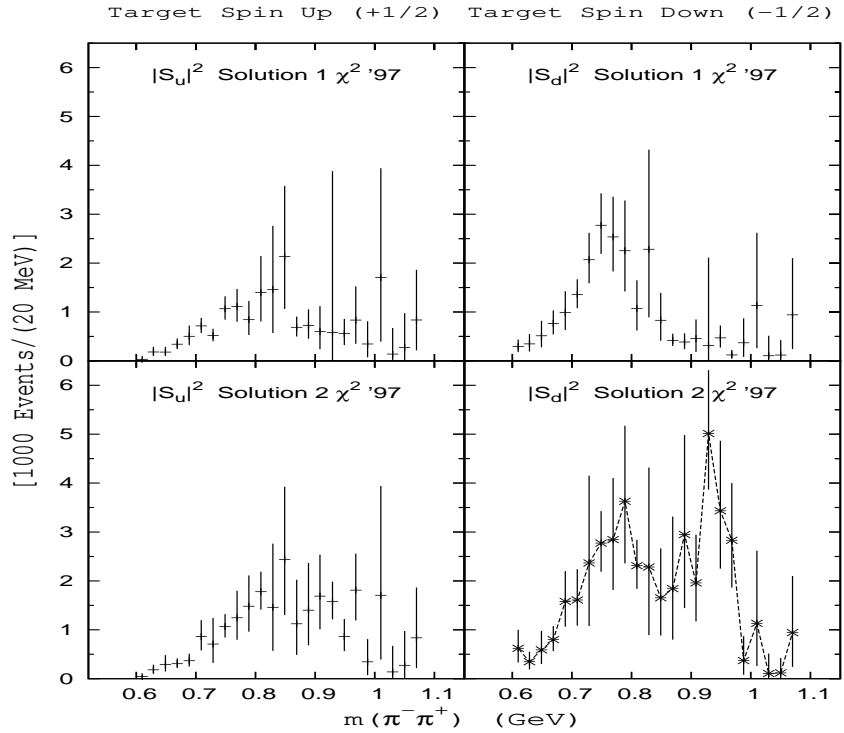}
\caption{Moduli of $S$-wave amplitudes $|S_u|^2$ and $|S_d|^2$ from the $\chi^2$ fit analysis~\cite{kaminski97}.}
\label{Figure 2}
\end{figure}

\begin{figure} [htp]
\includegraphics[width=12cm,height=10.5cm]{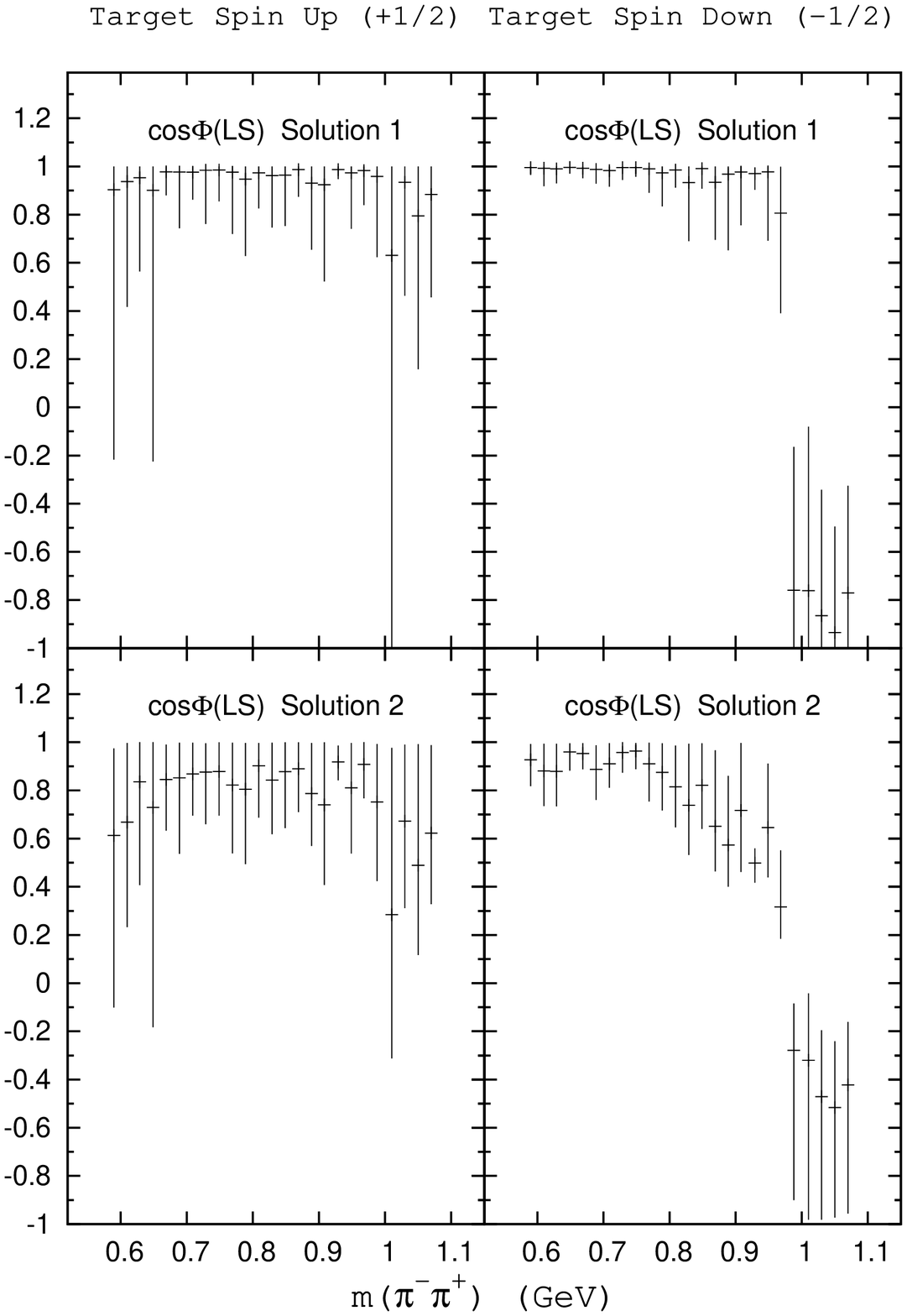}
\caption{Cosines $\cos \Phi_\tau(LS)$ from the Monte Carlo amplitude analysis~\cite{svec12a}.}
\label{Figure 3}
\end{figure}

\begin{figure} [hp]
\includegraphics[width=12cm,height=10.5cm]{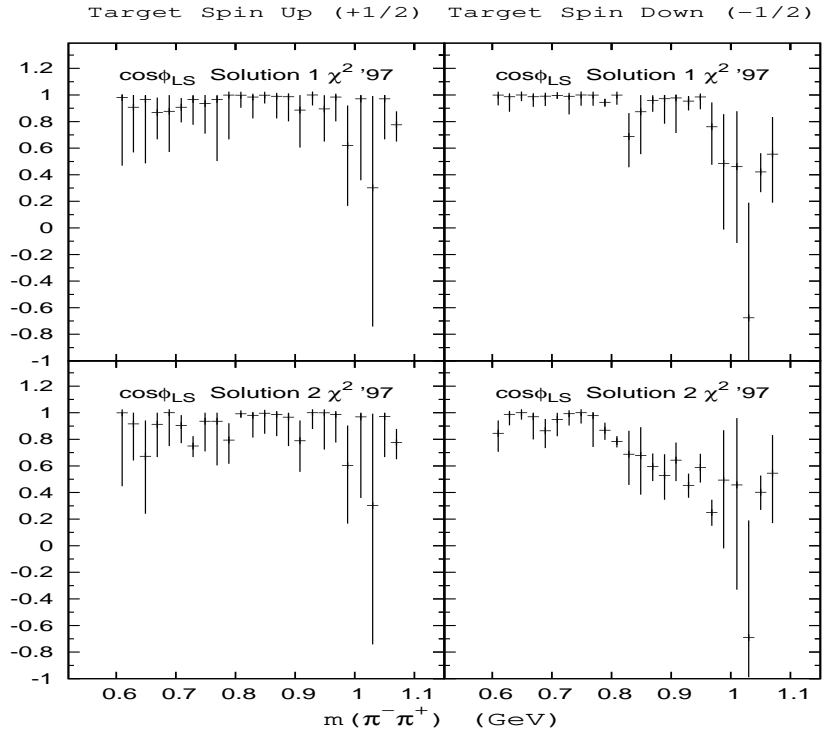}
\caption{Cosines $\cos \Phi_\tau(LS)$ from the $\chi^2$ fit amplitude analysis~\cite{kaminski97}.}
\label{Figure 4}
\end{figure}

\begin{figure} [htp]
\includegraphics[width=12cm,height=10.5cm]{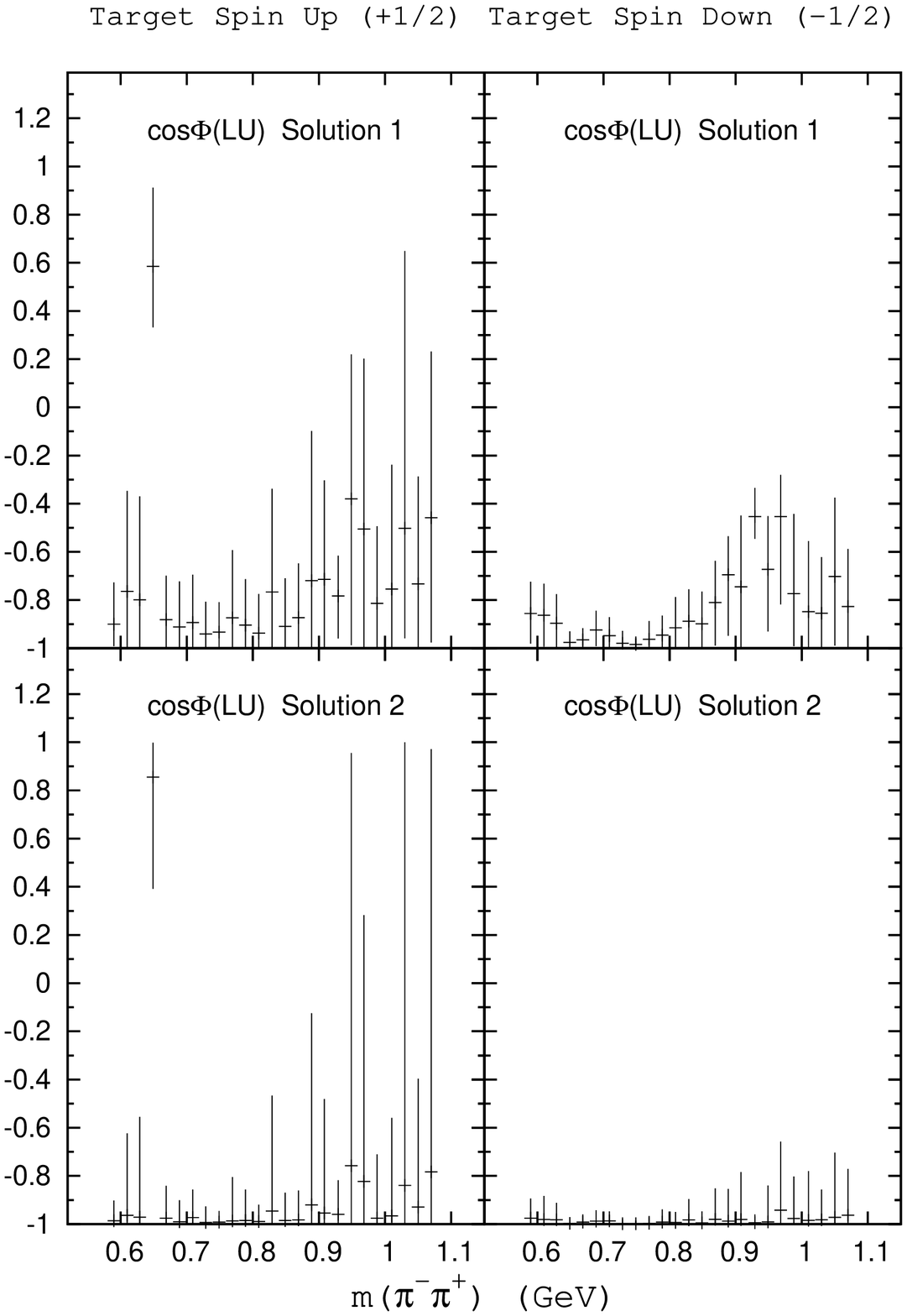}
\caption{Cosines $\cos \Phi_\tau(LU)$ from the Monte Carlo amplitude analysis~\cite{svec12a}.}
\label{Figure 5}
\end{figure}

\begin{figure} [hp]
\includegraphics[width=12cm,height=10.5cm]{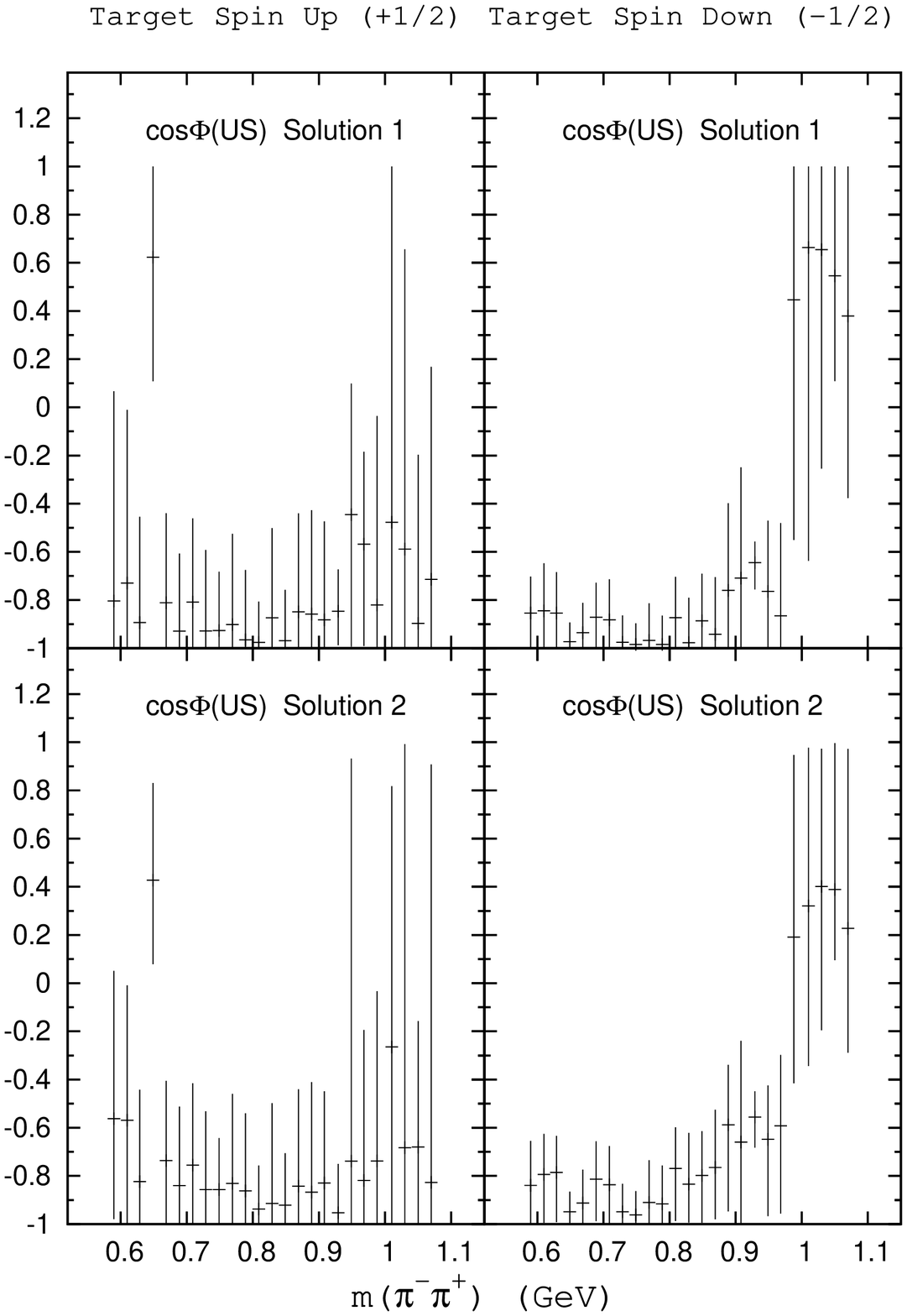}
\caption{Cosines $\cos \Phi_\tau(US)$ from the Monte Carlo amplitude analysis~\cite{svec12a}.}
\label{Figure 6}
\end{figure}

\subsubsection{Predictions for the cosines of the relative phases}

Figures 3 and 4 show the cosines for the relative phases $\Phi_\tau(LS)$ for the Monte Carlo~\cite{svec12a} and for the $\chi^2$ fit~\cite{kaminski97} analyses of the same CCM data, respectively. The results are again nearly identical. The most important but subtle difference is in the cosines for $\tau=d$ in the mass interval 700-800 MeV. While in the Monte Carlo analysis the values of this cosine are always less than 1.0, in the $\chi^2$ fit analysis this cosine takes on the value of 1.0 at 749 MeV bin. This difference means that in the Monte Carlo analysis there can be no change of sign of the phase $\Phi_\tau(LS)$ for $\tau=d$ below 1080 MeV while this phase can change sign at the 749 MeV bin in the $\chi^2$ analysis.

Figures 5 and 6 show the cosines for the relative phases $\Phi_\tau(LU)$ and $\Phi_\tau(US)$, respectively, for the Monte Carlo analysis~\cite{svec12a}. There are no corresponding Figures available from the $\chi^2$ fit analysis~\cite{kaminski97}.

A careful inspection of the Figures 3, 5 and 6 shows that with the exception of Solution 2 for $\cos \Phi_\tau(LS)$ the cosines of relative phases in Monte Carlo amplitude analysis are near the unitary values (8.10) and (8.11) as expected from the spin mixing relations (8.18) and (8.19).  Also with the exception of Solution 2 for $\cos \Phi_\tau(LS)$, the cosines are approximately independent of the transversity $\tau$. This consistency is approximate because the amplitude analysis did not impose the spin mixing mechanism constraints $|S_u|=K|S_d|$ and $|L_u|=K|L_d|$. When these constraints are imposed in the amplitude analysis the consistency for $\cos \Phi_\tau(LS)$ becomes exact~\cite{svec14a}.

\section{The dimension $M$ of the Hilbert space $H(E)$.}

We now turn to the question of the dimension $M$ of the Hilbert space $H(E)$. Recall that $M=\dim H(E) \leq \dim H(S_i) \dim H(S_f)$ in a non-unitary process $\rho(S_i) \to \rho(S_f)$~\cite{nielsen00}. In the Kraus representation (3.7) $\rho(S_i)=\rho_f(S)$ and $\rho(S_f)=\rho_f(O)$. The relevant Hilbert space $H(O)=H(S)$ is the space of spin states for all final state particles forming the states $\rho_f(S)$ and $\rho_f(O)$. The dimension $M$ must be common to all scattering processes. 

The simplest scattering process that involves interaction with the environment is $\pi N \to \pi \pi N$. At dipion mass $m$ the dimension $N(m)=\dim H(O)= \dim H(S)$ is determined by the maximum dipion spin $J_{max}(m)$. Its simplest spin state describes the $S$-wave process $\pi N \to 0^{++} N$. In the absence of $P$-waves in processes like $\pi^-p \to \pi^0 \pi^0 n$ or $\pi^+ p \to \pi^+ \pi^+ n$ this is the only process near threshold. The BNL high statistics meaurements of $\pi^-p \to \pi^0 \pi^0 n$ at 18.2 GeV/c show that all $D$-wave amplitudes are consistent with zero for $m \lesssim 400$ MeV for all momentum transfers $t$ \cite{gunter01}. The evolution of $0^{++} N \to 0^{++}N$ due to its interaction with the environment then limits the dimension $M$ to $1 \leq M \leq (2{1\over{2}}+1)(2{1\over{2}}+1)=4$. 

Experimental evidence for $M>1$ can only come from the violation of the cosine condition (3.11). Consider three observed amplitudes $|A|$, $|B|$ and $|C|$ and their correlations
$\cos \alpha=\cos \Phi_{AB}$, $\cos \beta= \cos \Phi_{AC}$ and 
$\cos \gamma=\cos \Phi_{BC}$. In general the phases  $\alpha$, $\beta$ and $\gamma$ violate the phase condition (3.10) but satisfy a phase gap condition
\begin{equation}
\alpha-\beta-\gamma=\Delta
\end{equation}
which leads to a cosine gap condition
\begin{equation}
\cos^2 \alpha +\cos^2 \beta + \cos^2 \gamma -2\cos \alpha \cos \beta \cos \gamma = 1 +G \equiv \Gamma
\end{equation}
where $\Delta$ and $G$ are the phase gap and cosine gap, respectively. They are related by the condition
\begin{equation}
G=\Bigl ( \sin \beta \sin \gamma +2\sin {\Delta \over{2}} \sin (\beta + \gamma + {\Delta \over{2}}) \Bigr )^2 - \Bigl ( \sin \beta \sin \gamma \Bigr )^2 
\end{equation}
The cosine gap condition (9.2) is an unambigous signature of the non-unitary evolution law with $M>1$.

The first indication for the violation of the phase condition (9.1) and the cosine condition (9.2) came in 1985 from the amplitude analysis of CERN measurement of $\pi^- p \to \pi^- \pi^+ n$ at 17.2 GeV/c at large momentum transfers $0.2 \leq |t| \leq 1.0$ (GeV/c$)^2$ by Rybicki and Sakrejda~\cite{rybicki85}. In that analysis they determined relative phases of $P$-wave amplitude $P^0_\tau$ ($\equiv L_\tau$ in our notation) and $D$-wave amplitudes $D^{0}_\tau$ and $D^{2U}_\tau$. Figure 10 shows the phase gap
\begin{equation}
\Delta_\tau =-\Phi(P^0_\tau D^{0}_\tau) + \Phi(P^0_\tau D^{2U}_\tau)+\Phi(D^{2U}_\tau D^{0}_\tau)
\end{equation}
and the cosine gap
\begin{equation}
\Gamma_\tau=\cos^2\Phi(P^0_\tau D^{0}_\tau)+\cos^2\Phi(P^0_\tau D^{2U}_\tau)+
\cos^2\Phi(D^{2U}_\tau D^{0}_\tau)
\end{equation}
\[
-2\cos\Phi(P^0_\tau D^{0}_\tau) \cos\Phi(P^0_\tau D^{2U}_\tau)
\cos\Phi(D^{2U}_\tau D^{0}_\tau)
\]
These phase gaps and cosine gaps show large deviations from the unitary values 0 and 1, respectively, as well as large fluctuations. The analysis of Rybicki and Sakrejda supports the non-unitary evolution law with $M>1$. Their analysis also finds evidence for a rho-like resonance in the $S$-wave, suggesting $\rho^0(770)-f_0(980)$ mixing at large $t$ (see also Ref.~\cite{svec12d}).   

\begin{figure} [hp]
\includegraphics[width=12cm,height=10.5cm]{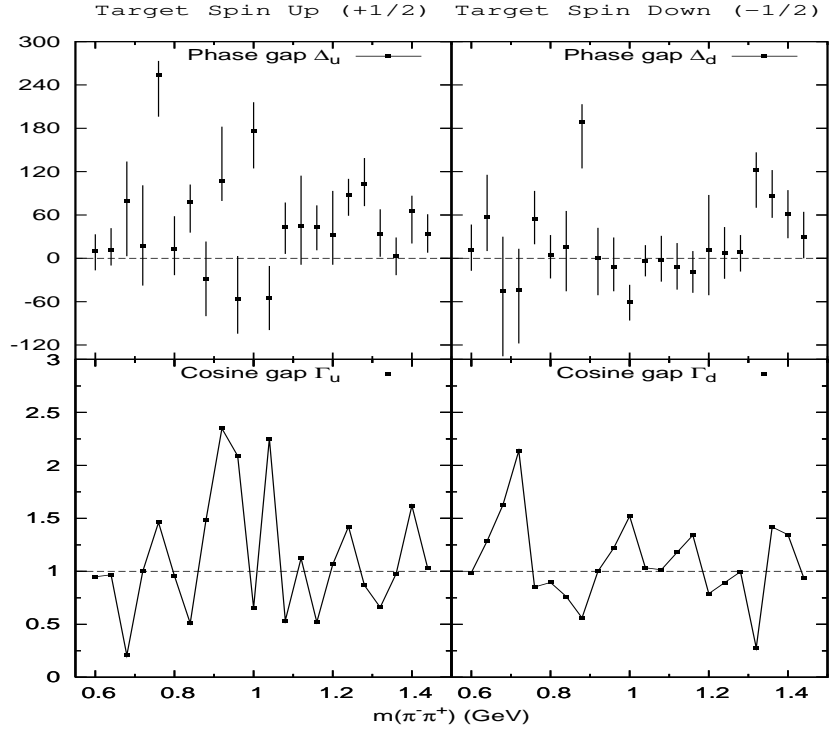}
\caption{Phase gap $\Delta_\tau$ and cosine gap $\Gamma_\tau$ for aplitudes $P^0_\tau$, $D^0_\tau$ and $D^{2U}_\tau$. Data from Ref.~\cite{rybicki85}.}
\label{Figure 7.}
\end{figure}

In our new analysis~\cite{svec14a} the spin mixing mechanism allows to extract $D$-wave observables from the CERN data. Analysis of the full $D$-wave subsystem for transversity $\tau=u$ reveals a violation of cosine conditions (3.11) by the amplitudes $D^{2U}_u$ and $D^{2N}_u$. Detailed analysis of the amplitudes $D^{2U}_u$ and $D^{2N}_u$ determines that the number of interacting degrees of freedom of the environment is $M=4$.

\section{Decoherence free and decohering amplitudes below 1400 MeV.}

The evidence that $M=\dim H(E)>1$ brings up the issue of decoherence free and decohering amplitudes. This issue is relevant since the decoherence free subspaces are linked to observations of spin mixing.

Kraus amplitudes forming decoherence free subspace do not depend on the interacting degrees of freedom $\ell$ since all Kraus operators are equal $V_\ell=V_0$, $ \ell=1,M$. In contrast the decohering Kraus amplitudes do depend on $\ell$ due to different Kraus operators. Three or more Kraus amplitudes are recognized as decoherence free amplitudes if all cosine conditions on the cosines derived from their measured bilinear terms are satisfied. Then we refer to the measured amplitudes also as decoherence free amplitudes. If one or more cosine conditions are violated they evidence one or more decohering Kraus amplitudes and we refer to the measured amplitudes as decohering amplitudes. In the absence of a theory of dephasing interactions the spin mixing matrices $V^\lambda_{JK}(\ell)$ are only constrained by the experimental data from which we can deduce which amplitudes are decoherence free and which are decohering.

In the sequel paper~\cite{svec14a} we report a new amplitude analysis of the CERN data below 1080 MeV using spin mixing mechanism (8.20) and (8.22) and the constraints (8.29) to determine anew the spin mixing $S$-and $P$-wave transversity amplitudes $S_\tau, L_\tau$, and to determine the $S$-matrix transversity amplitudes $S^0_\tau,L^0_\tau$ and the spin mixing parameters $\theta, \phi$. The spin mixing mechanism also allows to extract full set of $D$-wave amplitudes for the transversity $\tau=u$ in a simultaneous analysis. The spin mixing mechanism admits only a single physical solution for the $S$-and $P$-wave amplitudes and for the $D$-wave amplitudes $D^0_u,D^U_u,D^N_u$ corresponding to the Solution 2 in Figures 1-6. Interpreting this solution as Kraus amplitudes implies that these amplitudes form a decoherence free subspace because $M>1$. The spin mixing parameters $\theta, \phi$ do not depend on $\ell$.  

\begin{figure} [hp]
\includegraphics[width=12cm,height=10.5cm]{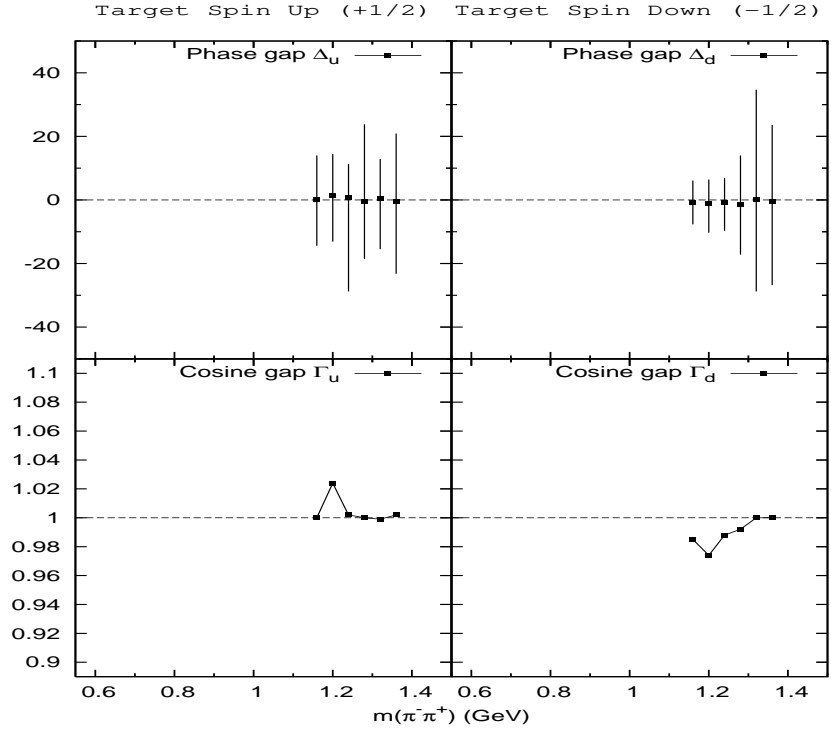}
\caption{Phase gap $\Delta_\tau$ and cosine gap $\Gamma_\tau$ for aplitudes $P^0_\tau$, $S_\tau$ and $D^0_\tau$. Data from Ref.~\cite{becker79b}.}
\label{Figure 8.}
\end{figure}

There is no evidence for the presence of $\rho^0(770)$ in the amplitude $D^0$ in our analysis~\cite{svec14a} and in the high $t$ analysis~\cite{rybicki85}, as predicted by the equations (8.5). We do not expect $f_2(1270)$ to mix with $L_\tau$ because $D^0_\tau$ still decouples from the $S$- and $L$-wave amplitudes in agreement with the data. Figure 11 shows that in the analysis at low $t$ and $m>980$ MeV~\cite{becker79b} the phase condition and cosine condition are satisfied by the relative phases $\Phi_\tau(LS)$, $\Phi_\tau(LD^0)$ and $\Phi_\tau(D^0S)$ for both $\tau$.

The analysis of the full $D$-wave subsystem reveals $\rho^0(770)$ mixing in the amplitudes $|D^U_u|^2$ and $|D^N_u|^2$ which is at variance with the analysis at high $t$~\cite{rybicki85}. This mixing requires a small mixing of the $S$-matrix amplitudes $D^{U,0}_u$ and $D^{N,0}_u$ in the amplitudes $U_u$ and $N_u$ in (8.2) and (8.3), respectively 
\begin{eqnarray}
U_u & = & V^1_{11}U^0_u+V^1_{12}D^{U,0}_u=e^{i2\phi}U^{0,eff}_u \approx e^{i2\phi}U^0_u\\
\nonumber
N_u & = & V^1_{11}N^0_u+V^1_{12}D^{N,0}_u=e^{i2\phi}N^{0,eff}_u \approx e^{i2\phi}N^0_u
\end{eqnarray}
where the mixing matrix $V^1_{JK}$ has a general form (8.13). The effective amplitudes $U^{0,eff}_u$ and $N^{0,eff}_u$ have the same $S$-matrix phases as $U^0_u$ and $N^0_u$. The amplitudes $D^{U,0}_u$ and $D^{N,0}_u$ have the same phases as $U^0_u$ and $N^0_u$, respectively~\cite{svec13a}. Amplitude analysis determines the effective amplitudes. Above 980 MeV there is no evidence for the presence of $f_2(1270)$ in the amplitudes $U_\tau$ and $N_\tau$~\cite{becker79b,chabaud83,rybicki85} indicating $V^1_{12} \approx 0$ at these masses. 

Our analysis of the amplitudes $D^{2U}_\tau$ and $D^{2N}_\tau$ in Ref.~\cite{svec14a} determines four phases $\chi(\ell),\ell=1,4$ for the two decohering amplitudes $D^{2U}_\tau(\ell)$ and $D^{2N}_\tau(\ell)$ in (8.4) 
\begin{equation}
V^2_{22}(\ell)=e^{i\chi(\ell)}
\end{equation}
which sets the dimension $M=4$. The results for the $D$-wave amplitudes mean that the amplitudes $D^0_\tau$, $D^U_\tau$ and $D^N_\tau$ belong to the same decoherence free subspace with the $S$-and $P$-wave amplitudes. This subspace includes three pairs of spin mixing amplitudes ($S_\tau,L_\tau$), ($U_\tau,D^U_\tau$) and ($N_\tau,D^N_\tau$). We conclude with the observation that the spin mixing occurs in decoherence free subspaces while there is no spin mixing in decohering subspaces. 
 
\section{Spin mixing mechanism and the violation of Lorentz symmetry.}

We shall use the Coleman-Mandula No-Go Theorem~\cite{coleman67} to show that the non-unitary evolution law (3.7) is consistent with the $S$-matrix scattering dynamics and quantum field theory provided the dimension $M>1$. The Theorem requires that in the decohering subspaces there can be no violation of Lorentz symmetry while there must be a Lorentz symmetry violating spin mixing in the decoherence free subspaces.

A modern proof of the Theorem is given by Weinberg in Ref.~\cite{weinberg00}.
Assume that the generators $P_\mu$ and $M_{\mu \nu}$ of the Poincare group commute with the $S$-matrix. Assume that internal symmetry generators $T^a$ (charges) commute with the $S$-matrix and form an algebra given by commutation relations
\begin{equation}
[T^a,T^b]=if^{abc}T^c
\end{equation}
Coleman-Mandula No-Go Theorem asserts that there is no physically meaningful $S$-matrix theory where the charges $T^a$ do not commute with the generators of the Poincare group. That means there is no mixing of space-time transformations and internal symmetry transformations. Supersymmetry evades this No-Go Theorem by replacing the commutation relations (11.1) with anticommutation relations~\cite{weinberg00}. 

The key assumption in the proof of the Coleman-Mandula Theorem is the assumption that a two-body scattering always exists in the $S$-matrix theory as opposed to a situation where only a forward scattering is possible. Theories that allow only forward scattering evade the No-Go Theorem. The matrix elements of the operators $V_k$ describing the pure dephasing interaction of the particle final states of $\rho_f(S)$ all describe forward scattering of these states. Do they evade the No-Go Theorem?

To answer this question let us first assume that there is only one Kraus operator $V$. This operator $V$ cannot be the $S$-matrix for the $S$-matrix would allow a non-trivial scattering of the produced particle states. Thus $V$ is a genuine "No-scattering" operator. The Lorentz symmetry is conserved by $V$ and the Coleman-Mandula Theorem is evaded. Then conserved internal charges associated with the Kraus operator $V$ need not commute with the generators of the Poincare group and in fact can carry a Lorentz index. These charges would include isospin of particles involved in the forward scattering in a complete contradiction with the $S$-matrix theory including supersymmetry. To resolve this contradiction we note that the derivation of the No-Go Theorem assumes that the evolution of the scattering process is unitary. By assuming that there is more than one dephasing "No-scattering" operator in (3.7), i.e. by assuming a non-unitary evolution with $V_\ell,\ell=1,M>1$, we can "evade the evasion" and avoid any charges with Lorentz index to resolve the contradiction. 

The non-unitary evolution law does allow for unitary decoherence free channels in the Hilbert space $H(O)$. In these subspaces all Kraus operators $V_\ell$ are equal to a unitary operator $V_0$. In these spaces we cannot invoke the condition $M>1$ to "evade the evasion" of the No-Go Theorem. Neither can we invoke the breaking of the Lorentz symmetry by the Kraus operators. The only option the dynamical system now has to "evade the evasion" in order to preserve its consistency with the $S$-matrix dynamics is to break Lorentz symmetry by mixing some $S$-matrix amplitudes with different spins in the observed amplitudes. This is a classical case of a spontaneously broken symmetry induced by the environment. 

The spin mixing of the $S$-matrix amplitudes is a pure quantum mechanical effect due to the superposition principle. It arises from the action of the Kraus operators on the dipion-recoil nucleon states $|p_s p_d;J \lambda I_J, \chi> $ given by the superposition (4.14). Since it affects only the observed amplitudes it co-exists with the fundamental Lorentz symmetry of the particle dynamics and Standard Model.

Quantum superposition principle co-exists with Lorentz symmetry in another important quantum effect arising from this principle - quantum entanglement of spin states or other quantum states. The apparent violation of Lorentz symmetry by two entangled particles is well known since the famous EPR paper. In this sense the spin mixing in particle scattering is analogous to quantum entanglement. In both cases there is an apparent violation of the Lorentz symmetry.

Vector-scalar meson mixing is not a new idea. In 1977 Chin studied $\sigma(500)-\omega(783)$ mixing in nucleon-nucleon interactions in high density matter~\cite{chin77}. In 2000-2002 Gale and collaborators examined the effects of $\rho(770)-a_0(980)$ mixing on the dilepton production in relativistic heavy ion collisions~\cite{gale00,gale01,gale02a,gale02b} measured at RHIC. The $\sigma(500)-\omega(783)$ mixing was induced by the ground state of the system of the interacting nucleons, while the $\rho(770)-a_0(980)$ mixing originated in nucleon-nucleon excitations in the medium of nuclear matter. In both cases there was no violation of fundamental symmetries because the interaction Lagrangian conserved those symmetries including the Lorentz symmetry. The meson spin mixing and the spontaneous violation of Lorentz symmetry was due to the interaction of the nucleon-nucleon scattering process with its environment. It was this work on vector-scalar mixing that inspired the present author in 2003 to seek the solution to the puzzle of the appearance of a rho-like resonance in the $S$-wave in $\pi^- p \to \pi^- \pi^+ n$ and its absence in $\pi^- p \to \pi^0 \pi^0 n$ in the idea of $\rho^0(770)-f_0(980)$ spin mixing.

%\newpage
\section{Conclusions and outlook.}

The CERN measurements of $\pi^- p \to \pi^- \pi^+ n$ on polarized target at 17.2 GeV/c are so far the only high statistics measurements of its kind, yet they revealed an interesting new physics beyond the Standard Model. In Part I. of this work we have shown that they reveal a nonunitary interaction of the produced final state with a quantum environment. To preserve the identity of the final state particles this new interaction must be a pure dephasing interaction. In this Part II. we required that the dephasing interaction be fully consistent with the Standard Model. From the complexity of the consistency requirements emerges the simplicity of the spin mixing mechanism (6.17) and (6.18), and its validation by the conservation of probability (7.21). From this consistency alone we have deduced that in $\pi N \to \pi \pi N$ processes the new interaction with the environment must be a dipion spin mixing interaction governed by the spin mixing mechanism. Applied to the $S$- and $P$-wave subsystem in $\pi^- p \to \pi^- \pi^+ n$ the theory predicts $\rho^0(770)-f_0(980)$ mixing in agreement with the evidence for such a mixing in all analyses of five experiments on polarized targets surveyed in Ref.~\cite{svec12d}. Due to the selection rule $K=J-1,J,J+1$ the theory correctly predicts the absence of $f_2(1270)$ mixing in $\pi^- p \to \pi^- \pi^+ n$. 

Since only $J=even$ are allowed in $\pi^- p \to \pi^0 \pi^0 n$ and in $\pi^+ p \to \pi^+ \pi^+ n$ there is no spin mixing in these processes, only a change of the phase of the $S$-matrix partial wave amplitudes. Consider u-channel processes such as $p \bar{p} \to \pi^- \pi^+ \pi^0$ and $p \bar{p} \to \pi^0 \pi^0 \pi^0$. In these reactions the two-pion systems $\pi^-\pi^+$ and $\pi^0 \pi^0$ are accompanied by a spin 0 particle. The conservation of the total angular momentum by the Kraus operators in $<J \lambda,0;p_c p_d|V_\ell|K \mu,0;p_c p_d>$ then requires $K=J$. Therefore there is no spin mixing also in these processes. 

In a sequel paper~\cite{svec14a} a new amplitude analysis of the CERN data using spin mixing mechanism determines spin mixing parameters $\theta$ and $\phi$, $S$-and $P$-wave $S$-matrix amplitudes and selects a single solution for the spin mixing amplitudes of a decoherence free $S$- and $P$-wave subsystem. This analysis allows to extract $D$-wave observables from the CERN data. Analysis of the full $D$-wave subsystem for transversity $\tau=u$ reveals a $\rho^0(770)$ mixing in the dephasing doublets $U_\tau,D^U_\tau$ and $N_\tau,D^N_\tau$. Detailed analysis of the amplitudes $D^{2U}_u$ and $D^{2N}_u$ determines that the number of interacting degrees of freedom of the environment is $M=4$. Evidence from the analysis at high $t$~\cite{rybicki85} limits the dimension $M$ of the Hilbert space of the environment to $2 \leq M \leq 4$. 

We use Coleman-Mandula No-Go Theorem to show that the consistency of the pure dephasing interactions with the Standard Model requires $M>1$. As a result decoherence free subsystems must spontaneously break Lorentz symmetry while the Lorentz symmetry is conserved by the decohering subspaces. The breaking of Lorentz symmetry or the violation of cosine conditions in the observed amplitudes are thus two unambigous signatures of dephasing interactions.

The obvious question arises - what is the physical nature of the quantum environment? The consistency of the quantum environment with the Standard Model suggests that it is a universal environment in the entire Universe. It can be viewed as a sea of quantum states $\rho(E)$ across the Universe. The quantum states $\rho(E)$ must be stable and cannot directly interact with any of the particles of the Standard Model including the Higgs boson. 

The pure dephasing interaction involves only the exchange of quantum information  between the two quantum states $\rho_f(S)$ and $\rho_i(E)$. As such it stands outside of the Standard Model. While the dephasing interaction can have strong observable effects, such as $\rho^0(770)-f_0(980)$ mixing, its effects are not observable in the usual studies of particle scattering. Since these experiments do not involve diparticle partial wave amplitudes in measurements on polarized targets, they cannot observe the resonance mixing or the violations of the cosine conditions. The measurements of relative phases alone cannot provide evidence for the quantum environment since the needed unitary phases are known only for $\pi N\to \pi \pi N$ processes. The dephasing interaction is largely a "dark" interaction. 

The universal presence of the quantum environment and its pure dephasing interaction with the baryonic matter must manifest themselves in astrophysical observations. Such observations provide a convincing evidence for the existence of dark matter and dark energy which are both omnipresent environments in the Universe with non-standard interactions with baryonic matter. In Ref.~\cite{svec14a} we present a physically motivated model of the quantum states $\rho(E)$ which we propose to identify with the particles of a distinct component of the cold dark matter. In this picture pure dephasing interactions are the interactions of baryonic matter with this form of dark matter. These interactions are not rare events but they require high statistics measurements on polarized targets for their detection. Dedicated measurements of production processes such as $\pi N \to \pi \pi N$ and $K N \to K \pi N$ on polarized targets would provide new and accessible tools to explore the dark sector of the Universe.


\newpage
\begin{thebibliography} {}

\bibitem{hagopian63} V.~Hagopian and W.~Selow, {\sl Experimental Evidence on $\pi \pi$ Scattering Near the $\rho^0$ and $f^0$ Resonances from $\pi^- + p \to \pi + \pi + nucleon$   at 3 BeV/c}, Phys.Rev.Lett. {\bf 10}, 533 (1963).

\bibitem{islam64} M.M.~Islam and R.~Pinn, {\sl Study of $\pi^- \pi^+$ System in $\pi^- + p \to \pi^- + \pi^+ + n$ Reaction}, Phys.Rev.Lett. {\bf 12}, 310 (1964).

\bibitem{patil64} S.H.~Patil, {\sl Analysis of the $S$-wave in $\pi \pi$ Interactions}, Phys.Rev.Lett. {\bf 13}, 261 (1964).

\bibitem{durand65} L.~Durand III and Y.T.~Chiu, {\sl Decay of the $\rho^0$ Meson and the Possible Existence of a $T=0$ Scalar Di-Pion}, Phys.Rev.Lett. {\bf 14}, 329 (1965).

\bibitem{baton65} J.P.~Baton {\sl et al.}, {\sl Single Pion Production in $\pi^-p$ Interactions at 2.75 GeV/c}, Nuovo Cimento {\bf 35}, 713 (1965).

\bibitem{donohue79} J.T.~Donohue and Y.~Leroyer, {\sl Is There a Narrow $\epsilon$ under $\rho^0$ ?}, Nucl.Phys. {\bf B158}, 123 (1979).

\bibitem{becker79a} H.~Becker {\sl et al.}, {\sl Measurement and Analysis of Reaction $\pi^- p \to \rho^0 n$ on Polarized Target}, Nucl.Phys. {\bf B150}, 301 (1979).

\bibitem{becker79b} H.~Becker {\sl et al.}, {\sl A Model Independent Partial-wave Analysis of the $\pi^+ \pi^-$ System Produced at Low Four-momentum Transfer in the Reaction $\pi^- p_{\uparrow} \to \pi^+ \pi^- n$ at 17.2 GeV/c}, Nucl.Phys. {\bf B151}, 46 (1979).

\bibitem{chabaud83} V.~Chabaud {\sl et al.}, {\sl Experimental Indications for a $2^{++}$ non-${\overline q}q$ Object},  Nucl.Phys. {\bf B223}, 1 (1983).

\bibitem{rybicki85} K.~Rybicki and I.~Sakrejda, {\sl Indication for a Broad $J^{PC}=2^{++}$ Meson at 840 MeV Produced in the Reaction $\pi^- p \to \pi^- \pi^+ n$ at High $|t|$}, Zeit.Phys. {\bf C28}, 65 (1985).

\bibitem{kaminski97} R.~Kami\'{n}ski, L.~Le\'{s}niak and K.~Rybicki, {\sl Separation of $S$-Wave Pseudoscalar and Pseudovector Amplitudes In $\pi^- p \to \pi^+\pi^- n$ On Polarized Target}, Zeit.Phys. {\bf C74}, 79 (1997).

\bibitem{kaminski02} R.~Kami\'{n}ski, L.~Le\'{s}niak and K.~Rybicki, {\sl A Joint Analysis of the $S$-wave in the $\pi^+\pi^-$ and $\pi^0\pi^0$ Data}, Eur.Phys.J.direct {\bf C4}, 1 (2002).

\bibitem{lesquen85} A.~de Lesquen, L.~van Rossum, M.~Svec {\sl et al.}, {\sl Measurement of the Reaction $\pi^+ n \to \pi^+ \pi^- p$ at 5.98 and 11.85 GeV/c Using a Transversely Polarized Deuteron Target}, Phys.Rev. {\bf D32}, 4355 (1985).

\bibitem{svec92c} M.~Svec, A.~de Lesquen and L.~van Rossum, {\sl Evidence for a Scalar State I=0 $0^{++}(750)$ from Measurements of $\pi N \to \pi^+ \pi^- N$ on a Polarized Target at 5.98, 11.85 and 17.2 GeV/c}, Phys.Rev. {\bf D46}, 949 (1992).

\bibitem{svec96} M.~Svec, {\sl Study of $\sigma(750)$ and $\rho^0(770)$ Production in Measurements of $\pi N \to \pi^+ \pi^- N$ on a Polarized Target at 5.98, 11.85 and 17.2 GeV/c}, Phys.Rev. {\bf D53}, 2343 (1996).

\bibitem{svec97a} M.~Svec, {\sl Mass and Width of the $\sigma(750)$ Scalar Meson from Measurements of $\pi N \to \pi^- \pi^+ N$ on Polarized Target}, Phys.Rev. {\bf D55}, 5727 (1997).

\bibitem{alekseev99} I.G.~Alekseev {\sl et al.} (ITEP Collaboration), {\sl Study of the Reaction $\pi^- p \to \pi^- \pi^+ n$ on the Polarized Proton Target at 1.78 GeV/c: Experiment and Amplitude Analysis}, Nucl.Phys. {\bf B541}, 3 (1999).

\bibitem{svec12d} M.~Svec, {\sl Consistency Tests of $\rho^0(770)-f_0(980)$ Mixing in $\pi^- p \to \pi^- \pi^+ n$}, arXiv:1411.2792 [hep-ph] (2014).

\bibitem{apel72} W.D.~Apel {\sl et al.}, {\sl Results on $\pi \pi$ Interaction in the Reaction $\pi^- p \to \pi^0 \pi^0 n$ at 8 GeV/c}, Phys.Lett. {\bf B41}, 542 (1972).

\bibitem{gunter01} J.~Gunter {\sl et al.} (BNL E852 Collaboration),
{\sl Partial Wave Analysis of the $\pi^0 \pi^0$ System Produced in $\pi^- p$ Charge Exchange Collisions}, Phys.Rev. {\bf D64}, 072003 (2001).

\bibitem{svec12a} M.~Svec, {\sl Evidence for $\rho^0(770)-f_0(980)$ Mixing in $\pi^- p \to \pi^- \pi^+ n$ from CERN Measurements on Polarized Target}, arXiv:1205.6391 [hep-ph] (2012).

\bibitem{svec13a} M.~Svec, {\sl Study of $\pi N \to \pi \pi N$ Processes on Polarized Targets I; Quantum Environment and Its Dephasing Interacton with Particle Scattering}, arXiv:1304.5120 [hep-ph] (2013).

\bibitem{svec14a} M.~Svec, {\sl Spin Mixing Mechanism in Amplitude Analysis of $\pi^- p \to \pi^- \pi^+ n$ and a New View of Dark Matter}, arXiv:1411.4468 [hep-ph] (2014).

\bibitem{kraus71} K.~Kraus, {\sl General State Changes in Quantum Theory}, Ann.Phys.(NY) {\bf 64}, 311 (1971).

\bibitem{kraus83} K.~Kraus, {\sl States, Effects, and Operations: Fundamental Notions of Quantum Theory}, Lecture Notes in Physics, Vol.190, Springer-Verlag, 1983.

\bibitem{nielsen00} M.A.~Nielsen and I.L.~Chuang, {\sl Quantum Computation and Quantum Information}, Cambridge University Press, 2000.

\bibitem{bengtsson06} I.~Bengtsson and K.~\.{Z}yczkowski, {\sl Geometry of Quantum States - An Introduction to Quantum Entanglement}, Cambridge University Press, 2006.

\bibitem{lidar99} D.A.~Lidar, D.~Bacon and K.B.~Whaley, {\sl Concatenating Decoherence Free Subspaces with Quantum Error Correcting Codes.}, Phys.Rev.Lett. {\bf 82}, 4556 (1999).

\bibitem{martin70} A.D.~Martin and T.D.~Spearman, {\sl Elementary Particle Theory}, North-Holland, 1970.

\bibitem{jacob59} M.~Jacob and G.C.~Wick, {\sl On the General Theory of Collisions for Particles with Spin}, Ann.Phys.(NY), {\bf 7}, 404 (1959).

\bibitem{wick62} G.C.~Wick, {\sl Angular Momentum States for Three Particles}, Ann.Phys.(NY), {\bf 18}, 65 (1962). 

\bibitem{seiden05} A.~Seiden, {\sl Particle Physics: A Comprehensive Introduction}, Addison Wesley, 2005.

\bibitem{coleman67} S.~Coleman and J.~Mandula, {\sl All Possible Symmetries of the $S$-Matrix}, Phys.Rev. {\bf 159}, 1251 (1967).

\bibitem{weinberg00} S.~Weinberg, {\sl The Quantum Theory of Fields, Volume III Supersymmetry}, Cambridge University Press, 2000.

\bibitem{chin77} S.A.~Chin, {\sl A Relativistic Many-Body Theory of High Density Matter}, Annals of Physics (NY) {\bf 108}, 301 (1977).

\bibitem{gale00} O.~Teodorescu, A.K.~Muzunder and Ch.~Gale, {\sl Matter Induced $\rho-\delta$ Mixing: A Source of Dileptons}, Phys.Rev. {\bf C61}, 051901 (2000). 

\bibitem{gale01} O.~Teodorescu, A.K.~Muzunder and Ch.~Gale, {\sl Effects of Meson Mixing on Dilepton Spectra}, Phys.Rev. {\bf C63}, 034903  (2001).

\bibitem{gale02a} O.~Teodorescu, A.K.~Muzunder and Ch.~Gale, {\sl Aspects of Meson Properties in Dense Nuclear Matter}, Phys.Rev. {\bf C66}, 015209 (2002).

\bibitem{gale02b} O.~Teodorescu, A.K.~Muzunder and Ch.~Gale, {\sl Meson Mixing and Dilepton Production in Heavy Ion Collisions}, AIP Conf.Proc. {\bf 549}, 369 (2002).

\end{thebibliography}
\end{document}